\documentclass[letterpaper, 10 pt, conference]{ieeeconf} 

\IEEEoverridecommandlockouts                              % This command is only needed if 
\usepackage{cite}
\usepackage{amsmath,amssymb,amsfonts}
\usepackage{algorithmic}
\usepackage{graphicx}
\usepackage{textcomp}
\usepackage{xcolor} 

\newcommand{\st}[1]{${#1}^{\textrm{st}}$}
\newcommand{\nd}[1]{${#1}^{\textrm{nd}}$}

\def\R{\mathop{\textrm{I}\mkern -3.5mu \textrm{R}}} 

\def\C{{\setbox0=\hbox{$\displaystyle{\textrm{C}}$}
		\hbox{\hbox to0pt{\kern 0.4\wd0\vrule height 0.95\ht0\hss}\box0}}}

\newcommand{\abs}[1]{\left|{#1}\right|} 

\def\cF{\mathcal{F}} 
 
\def\cT{\mathcal{T}}
 
\def\cA{\mathcal{A}}

\usepackage{color}
\usepackage{xcolor}
\usepackage{colortbl}
\newtheorem{remark}{Remark}

\usepackage{amsmath,amssymb}
\usepackage{rotating}
\usepackage{epstopdf}
\usepackage{tikz} 
\usepackage[siunitx]{circuitikz}
\usepackage[percent]{overpic}
\usepackage{graphicx}

\usepackage{float}
\usepackage{longtable}
\usepackage{diagbox}
\usepackage{graphicx}
\usepackage{subcaption}
\usepackage{xcolor}
\usepackage{pict2e}

 \usepackage{flushend}

\usepackage{epsfig}
\usepackage[normalem]{ulem}

\def\R{\mathop{\textrm{I}\mkern -3.5mu \textrm{R}}} 

\def\C{{\setbox0=\hbox{$\displaystyle{\textrm{C}}$}
		\hbox{\hbox to0pt{\kern 0.4\wd0\vrule height 0.95\ht0\hss}\box0}}}

\def\cF{\mathcal{F}} 
 
\def\cT{\mathcal{T}}

\usepackage{placeins}

\def\BibTeX{{\rm B\kern-.05em{\sc i\kern-.025em b}\kern-.08em
    T\kern-.1667em\lower.7ex\hbox{E}\kern-.125emX}}

\begin{document}

%The 23rd IEEE Vehicle Power and Propulsion (IEEE VPPC 2026) will be held 5-9 October 2026, in Lyon, France.

\title{{\tt \scriptsize IEEE Vehicle Power and Propulsion Conference (VPPC) --- 5-9 October 2026, Lyon, France} \\ \LARGE \bf 
Model-Free Detection and Accommodation of Sensor Faults  For a PEM Electrolyzer}
%{\footnotesize \textsuperscript{*}Note: Sub-titles are not captured for https://ieeexplore.ieee.org  and
%should not be used}
%\thanks{Identify applicable funding agency here. If none, delete this.}

%\author{\IEEEauthorblockN{1\textsuperscript{st} Meziane Ait Ziane}
%\IEEEauthorblockA{\textit{Université de Lorraine, GREEN} \\
%%\textit{name of organization (of Aff.)}\\
%F-54000 Nancy, France \\
% meziane.ait-ziane@univ-lorraine.fr}
%\and
%\IEEEauthorblockN{2\textsuperscript{nd}  Michel Zasadzinski}
%\IEEEauthorblockA{\textit{Université de Lorraine, CRAN, UMR CNRS 7039} \\
%%\textit{name of organization (of Aff.)}\\
%F-54000 Nancy, France \\
%michel.zasadzinski@univ-lorraine.fr}
%\and
%\IEEEauthorblockN{3\textsuperscript{rd} Cédric Join}
%\IEEEauthorblockA{\textit{Université de Lorraine, CRAN, UMR CNRS 7039} \\
%%\textit{name of organization (of Aff.)}\\
%F-54000 Nancy, France\\
%cedric.join@univ-lorraine.fr}
%\and
%\IEEEauthorblockN{4\textsuperscript{th} Michel Fliess}
%\IEEEauthorblockA{\textit{ Sorbonne Université, LJLL, UMR CNRS 7598} \\
%%\textit{name of organization (of Aff.)}\\
%75005 Paris, France \\
%michel.fliess@sorbonne-universite.fr}
%} 

\author{Meziane Ait Ziane$^{1}$, Michel Zasadzinski$^{2}$,  Cédric Join$^{2}$ and Michel Fliess$^{3}$ % <-this % stops a space
\thanks{*This work was not supported by any organization}% <-this % stops a space
\thanks{$^{1}$M. Ait Ziane is with Université de Lorraine, GREEN, F-54000 Nancy, France.
        {\tt\footnotesize meziane.ait-ziane@univ-lorraine.fr}}%
\thanks{$^{2}$M. Zasadzinski and C. Join are with Research Center for Automatic Control of Nancy (CRAN), UMR CNRS 7039, Université de Lorraine, F-54000 Nancy, France
        {\tt\footnotesize \{michel.zasadzinski,cedric.join\}@univ-lorraine.fr}}%
 \thanks{$^{3}$M. Fliess is with Laboratoire  Jacques-Louis Lions (LJLL), UMR CNRS 7598, Sorbonne Université, 75005 Paris, France
        {\tt\footnotesize michel.fliess@sorbonne-universite.fr }}%
}
\allowdisplaybreaks
\maketitle

\begin{abstract}
We investigate the detection and accommodation of sensor faults for a proton exchange membrane electrolyzer coupled to a DC/DC converter powered by renewable energy sources. The proposed method for detecting and accommodating the sensor fault is model-free and is based on the concept of {\em ultra-local} model that is becoming classic in control engineering. The existing literature on active control tolerant to sensor fault dedicated to this question shows that no previous work has addressed this topic. Our approach mitigates the effect of sensor fault on closed-loop behavior and guarantees the stability and performance of the overall system. Numerical simulations under variations in renewable energy sources validate our approach. 
\end{abstract}

%\begin{IEEEkeywords}
%PEM electrolyzer, sensor fault tolerant-control, model-free control. 
%\end{IEEEkeywords}

\section{INTRODUCTION}\label{sec_intro} 
%\color{red} 
Green hydrogen is playing an key part in reducing carbon emissions in the industrial and transport sectors. This hydrogen is mainly produced using water electrolysers powered by renewable energy sources. In this framework, Proton Exchange Membrane Water Electrolyzers (PEMWE) technology is widely used for hydrogen production due to its fast response to renewable energy sources variations and its ability to support high current densities of up to 2\,A/cm$^2$ \cite{FaP:20}.

Due to its complexity, the PEMWE system is subject to several types of faults: those occurring at the stack or cell level, such as membrane drying, crossover, membrane degradation, \ldots, as well as those appearing at the system level, including balance of plant (leaks in pumps, valves, sensors, etc.). In recent years, many studies have been devoted to the diagnosis of this system. A model-based diagnosis approach via bond graph is proposed in \cite{sood2022robust}. The detection and isolation of several stack faults as membrane flooding and drying, crossover, \ldots are made by evaluating residuals. A signal-based approach is proposed in \cite{shen2025fault} to detect stack short circuits and water shortages and to detect pump fault in \cite{aubras2021non}. However, these methods consider the open loop operation of the system.

A PEMWE operates at a voltage significantly lower than that provided by Renewable Energy Sources (RES), thereby necessitating a DC/DC buck converter as an interface \cite{GGVD:20,YGKPHV:24}. An Active Fault-Tolerant Control (AFTC) strategy has been proposed for a three-phase stacked interleaved buck converter supplying a PEM electrolyzer \cite{guo2022new}, where a gate fault in one phase is considered. The proposed strategy enables continued operation under a degraded two-phase mode. Nevertheless, AFTC strategies addressing sensor faults accommodation for PEMWE systems remain largely unexplored and is the purpose of this work.

%\newpage 

This paper proposes an active sensor fault-tolerant control strategy for a PEMWE system coupled with DC/DC converter supplied by RES. The proposed approach ensures closed-loop operation under RES voltage variations and is model-free, i.e. it does not rely on a mathematical modeling of the system for fault detection and accommodation.%, and ensures closed-loop operation under RES voltage variations. 

This paper is organized as follows. Section \ref{sec_PEWE} describes the PEMWE and DC/DC converter models which is used as a test bench for validation and states the problem of sensor AFTC. Section \ref{sec_SFA} presents the proposed model-free sensor fault accommodation approach, accounting for variations in the input voltage. Simulation results are provided in Section \ref{sec_Sim}, and conclusions are drawn in Section \ref{sec_Conc}.

\color{black} 
\section{ PEMWE MODELING AND CONTROL OBJECTIVE} \label{sec_PEWE}
\subsection{Modeling and control objective}\label{ssec_modeling}
%\textcolor{red}{A PEMWE operates at low voltage compared to the voltage supplied by renewable energy sources. This configuration requires the integration of a DC/DC buck converter as an interface between the supplied energy source and the PEMWE \cite{GGVD:20,YGKPHV:24}.} The modeling of the entire system (PEMWE + SIBC) for green hydrogen production must take into account the PEMWE and the SIBC converter models.  

The PEMWE model used in this paper is the one given in \cite{ZAA:25} which has been identified using Electrochemical Impedance Spectroscopy (EIS) method detailed in \cite{VMZSKP:21,MSBSV:23}. The best obtained model in \cite{ZAA:25} is the one of order 6  and is given by \eqref{eq_EIS_models6}.
\begin{figure*}[h!]
 \centering 
 \rule{\linewidth}{0.5pt}
\begin{align}
    G(s) = \dfrac{ 0.02737 s^6 + 403 s^5 + 4.667e05 s^4 + 5.015e07 s^3 + 5.068e08 s^2 + 6.438e08 s + 1.009e08}{ s^6 + 1.051e04 s^5 + 8.42e06 s^4 + 6.784e08 s^3 + 4.788e09 s^2 + 4.214e09 s + 4.684e08}
    \label{eq_EIS_models6} 
\end{align}
 \rule{\linewidth}{0.5pt}
\end{figure*}

In order to obtain a model for the whole PEMWE + SIBC converter, we need to express the transfer function in \eqref{eq_EIS_models6} with passive electrical components. This can be achieved using the residue theorem \cite{Oga:70} as follows
\begin{align}
    G(s)=\sum_{h=1}^6\dfrac{R_h}{C_hR_hs+1}+R_0,
    \label{eq_EIS_models6_RC} 
\end{align}
where 
{\small
\begin{align*}
R_6&=100.06 \,\Omega, \ R_5=13.588\,\Omega, \ R_4=1.4068\,\Omega, \\
R_3&=0.20564 \,\Omega, \ R_2=0.040826 \,\Omega, \ R_1=0.0085339 \,\Omega, \\
C_6&=1.0361\times10^{-6}\,\textrm{F},\ C_5=9.4843\times10^{-5}\,\textrm{F}, \\ 
C_4&=0.0086629\,\textrm{F}, \, C_3=0.72622\,\textrm{F}, 
\ C_2=27.953\,\textrm{F}, \\
C_1&=901.43\,\textrm{F}, \, R_0=0.027366 \,\Omega.
\end{align*}}

\begin{figure*}
%\color{blue}
\centering
%\begin{center}
\begin{circuitikz}[american]
%\tikzstyle{estun}=[->,>=stealth]
\tikzstyle{estun}=[->,>=latex]
\tikzstyle{estun-b}=[thick]
\tikzstyle{estun-c}=[thick,dashed]
\draw[]%[color=black, thick]
% Electrolyzer
(0,0-0.5) to [short,-] (8+1,0-0.5){} % (0,0-0.5) to [short,-] (5.5,0-0.5){}
(0,2) to [short,-] (0,2){}
to[short, -, i=$I$] (1,2) ;
\draw (1,1.25) -- (1,2.75) ;
\draw[] %(1,2.75) -- (1.5,2.75) ;
(1,2.75) to [R,l=$R_1$,-] (3,2.75) ;
\draw (1,1.25) -- (1.5,1.25) 
(1,1.25) to [C,l=$C_1$,-] (3,1.25) ;
\draw (2.5,1.25) -- (3,1.25) ;
\draw (3,1.25) -- (3,2.75) ;
\draw[estun-c] (3,2) -- (3+1,2) ; %%%%%%%%%%%%%%%%%%
\draw[estun-c] (3+1,2) -- (3.5+1,2) ; %%%%%%%%%%%%%%%%%%
\draw (3.5+1,1.25) -- (3.5+1,2.75) ; %%%%%%%%%%%%%%%%%%
\draw[] %%%%%%%%%%%%%%%%%%
(3.5+1,2.75) to [R,l=$R_6$,-] (5.5+1,2.75) ; %%%%%%%%%%%%%%%%%%
\draw (5.5+1,1.25) -- (5.5+1,2.75) ; %%%%%%%%%%%%%%%%%%
\draw[] %%%%%%%%%%%%%%%%%%
(3.5+1,1.25) to [C,l=$C_6$,-] (5.5+1,1.25) ; %%%%%%%%%%%%%%%%%%
\draw[]
(5.5+1,2) to [R,l=$R_0$,-] (8+1,2) ;  % (3,2) to [R,l=$R_0$,-] (5.5,2) ; 
\draw (8+1,0) -- (8+1,2);
%\draw (8,0) to[V=$V_{\textrm{int}}$,invert] (8,2); % \draw (5.5,0) to[V=$V_{\textrm{int}}$,invert] (5.5,2) ;
\draw (8+1,0) -- (8+1,0-0.5); % \draw (5.5,0) -- (5.5,0-0.5);
%\draw[estun] ((0,0.15) -- (0,1.85) ;
%\draw[-triangle 45] (0,0.15) -- (0,1.85) ;
%\draw (0,0.2) -+ (0,1.4) node[vcc](VCC){} ;
\draw[->,>=latex] (0,0.15-0.5) -- (0,1.85) ;
\draw (0,1) node[right]{$V$} ;
%\draw (3,0.65) -+ (1,0.65) node[vcc](VCC){} ;
\draw[->,>=latex] (3,0.65) -- (1,0.65) ;
\draw (2,0.65) node[below]{$v_{c_1}$} ; %%%%%%%%%%%%%%%%%%
\draw[->,>=latex] (5.5+1,0.65) -- (3.5+1,0.65) ; %%%%%%%%%%%%%%%%%%
\draw (4.5+1,0.65) node[below]{$v_{c_6}$} ; %%%%%%%%%%%%%%%%%%
%\draw (0,2.2) -+ (0,3.4) node[vcc](VCC){} ;
\draw[->,>=latex] (0,2.15) -- (0,4) ;
\draw (0,3) node[right]{$v_{c_s}$} ;
% separateur
\draw[dotted,thick] (-0.2,-1.75) -- (-0.2,6) ; 
% DC-DC
\draw (0,0-0.5) -- (-8,0-0.5) ;
\draw (-0.75,0-0.5) -- (-0.75,0.75) 
(-0.75,0.75) to [C,l=$C_p$,-] (-0.75,1.25) ;
\draw (-0.75,1.25) -- (-0.75,2.75) 
(-0.75,2.75) to [C,l=$C_s$,-] (-0.75,3.25) ;
\draw (-0.75,3.25) -- (-0.75,4) ;
\draw[]
(-6.5,4) to[short,*-] (-4.5,4)
(-4.5,4) to[R=$R_{\ell_s}$] (-2.8,4) 
(-2.8,4) to[short, -, L=$L_s$] (-1.55,4)
(-1.55,4) to[short,-,i=$i_s$] (-0.75,4) ;
\draw[]
(-4.75,2) to[short,*-] (-4.5,2)
(-4.5,2) to[R=$R_{\ell_p}$] (-2.8,2) 
(-2.8,2) to[short, -, L=$L_p$] (-1.55,2)
(-1.55,2) to[short,-*,i=$i_p$] (-0.75,2) 
(-0.75,2) to[short,-,] (0,2) ;
\draw (-4.75,0-0.5) -- (-4.75,0.5) ;
\draw (-4.75,0-0.5) node[circ]{} ; % rond noir 
\draw (-4.75,1) node[nigbt,bodydiode](npn){}%{2}
%++(2,0)node[pigbt,bodydiode](npn){}
;
\draw (-5.15,0.35) node[]{$\cT_4$} ;
\draw (-4.75,1.5) -- (-4.75,4.5) ;
\draw %(-4.75,5) node[nigbt,bodydiode](npn){}%{2}
++(-4.75,5)node[pigbt,bodydiode](npn){}
;
\draw (-5.15,4.35) node[]{$\cT_3$} ;
\draw (-4.75,5.5) -- (-4.75,6) ;
\draw (-6.5,0-0.5) -- (-6.5,0.5) ;
\draw (-6.5,0-0.5) node[circ]{} ; % rond noir 
\draw (-6.5,1) node[nigbt,bodydiode](npn){}%{2}
%++(2,0)node[pigbt,bodydiode](npn){}
;
\draw (-6.9,0.35) node[]{$\cT_2$} ;
\draw (-6.5,1.5) -- (-6.5,4.5) ;
\draw %(-6.5,1) node[nigbt,bodydiode](npn){}%{2}
++(-6.5,5)node[pigbt,bodydiode](npn){}
;
\draw (-6.9,4.35) node[]{$\cT_1$} ;
\draw (-6.5,5.5) -- (-6.5,6) ;
\draw (-6.5,6) node[circ]{} ; % rond noir 
\draw (-4.75,6) -- (-8,6) ;
\draw[dashed] (-5.6,-0.25) -- (-5.6,-0.25-0.5) ;
%\draw[dashed] (-5.6,-0.25) -- (-5.6,1) ;
\draw[dashed] (-5.6,-0.25) -- (-5.6,-0.15) ;
\draw[dashed] (-5.6,{-0.15+sqrt(3)/4}) -- (-5.6,1) ;
%\draw[->,>=latex] (-5.6,1) -- (-5.2,1) ;
\draw (-5.6,1) node[circ]{} ; % rond noir
\draw[rotate=0,scale=1,thick] (-5.6+0.25+0,-0.15+0) -- (-5.6+0.25-0.5,-0.15+0) -- (-5.6+0.25-1/4,{-0.15+sqrt(3)/4}) -- cycle; % NOT
\draw[thick] (-5.6+0.25-1/4,{-0.15+sqrt(3)/4}) node[ocirc]{} ; % rond blanc
\draw[dashed] (-7.75,-0.25-0.5) -- (-7.75,1) ;
%\draw[->,>=latex] (-7.75,1) -- (-7,1) ;
\draw (-7.75,1) -- (-7,1) ;
\draw (-7.75,1) node[circ]{} ; % rond noir
%
%\draw[->,>=latex] (-7.3,5) -- (-7,5) ;
\draw[dashed] (-7.35,5) -- (-7.35,2) ;
\draw[dashed] (-7.35,2) -- (-5.6,2) ;
\draw[dashed] (-5.6,2) -- (-5.6,1) ;
%
%\draw[->,>=latex] (-5.6,5) -- (-5.2,5) ;
\draw[dashed] (-5.6,5) -- (-5.6,2.5) ;
\draw[dashed] (-5.6,2.5) -- (-7.75,2.5) ;
\draw[dashed] (-7.75,2.5) -- (-7.75,1) ;
\draw (-8,1) to[V=$V_{RES}$,invert] (-8,6);
\draw[dashed] (-7.75,-0.25-0.5) -- (-5.6,-0.25-0.5) ;
\draw (-8,0-0.5) -- (-8,1) ;
% PWM
\draw[->,>=latex] (-6.75,-0.75-0.5) -- (-6.75,-0.25-0.5) ;
\draw (-6.15,-0.75-0.5) rectangle (-7.35,-1.25-0.5) ; % -6.75
\draw (-6.1,-1-0.5) node[left]{PWM} ;
\draw[->,>=latex] (-6.75,-1.75-0.5) -- (-6.75,-1.25-0.5) ;
\draw (-6.75,-1.75-0.5) node[above right]{$u$} ;
\draw (-5.5,-1.5) node[above right]{Stacked interleaved} ;
\draw (-5.5,-2) node[above right]{DC-DC buck converter} ;
%\draw (0.5,-1.5) node[above right]{Model of PEM electrolyzer} ; 
\draw (0.5,-1.5) node[above right]{PEMWE equivalent electrical circuit} ;
\end{circuitikz}
%\end{center}
%\begin{figure}[ht]
%\centering
\vspace{-4mm}
\caption{Stacked interleaved DC-DC buck converter with pulse width modulation (PWM) and electrical model of PEMWE}
\label{fig-DCDC-Electrolyzer}
\end{figure*}
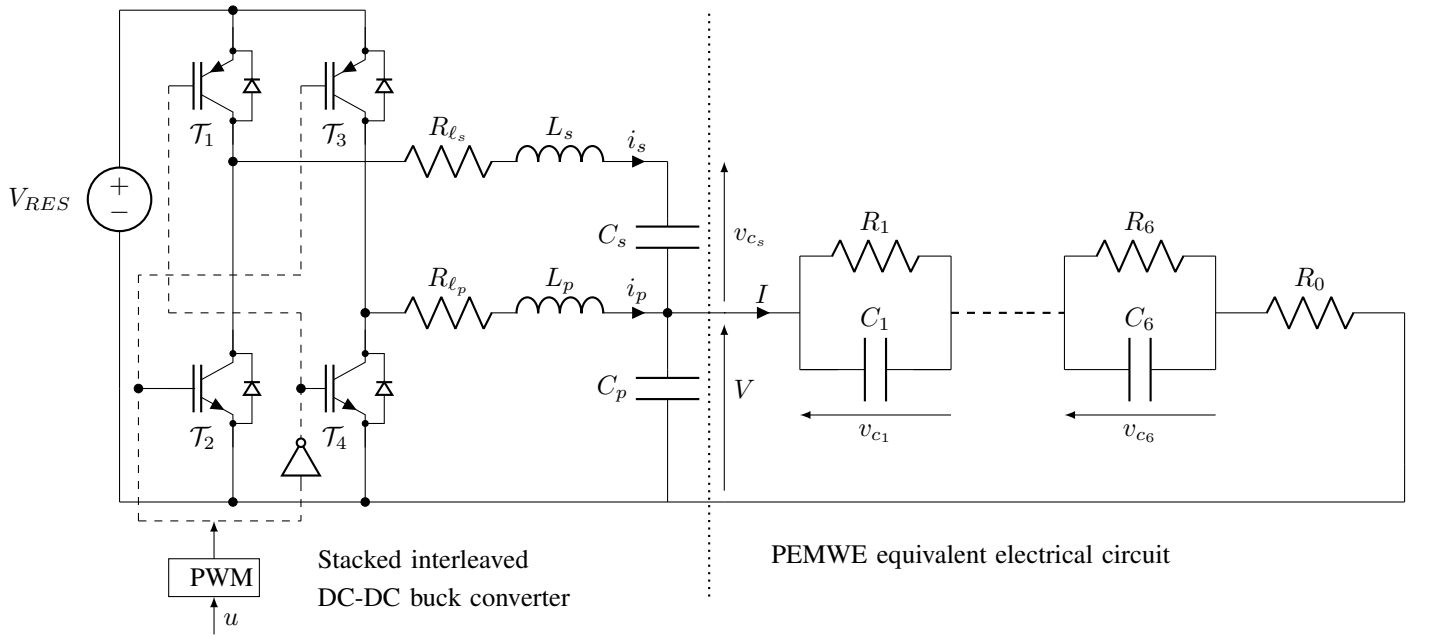

So the coupling of the PEMWE model given by \eqref{eq_EIS_models6_RC} with the SIBC converter is represented by Fig. \ref{fig-DCDC-Electrolyzer} where $V_{RES}$ is the voltage supplied by the RES, $u$ is the control input and the SIBC electrical components are given by: $L_p = L_s = 426 \times 10^{-6}\,$H, $R_{\ell_p} = R_{\ell_s} = 0.06\,\Omega$, $C_p = 10^{-4}\,$F and $C_s = 10^{-5}\,$F.

%Using the average state-space approach proposed in \cite{MiC:76} and detailed in \cite{MUR:95}, a state space model for the system given in Fig. \ref{fig-DCDC-Electrolyzer} is expressed by 

Using the average state-space approach proposed in \cite{MiC:76} and detailed in \cite{MUR:95}, a state-space model for the system given in Fig. \ref{fig-DCDC-Electrolyzer} is derived in \cite{MGZR:21} in the case of the number of parallel $RC$ cells is equal to 2 instead of 6 in \eqref{eq_EIS_models6_RC}. So using straightforward developments, the state-space model in \cite{MGZR:21} leads to the following one
\begin{subequations}\label{eq_sys2}
\begin{align}
     \dot{x}(t) &= Ax(t) + B V_{RES}(t) u(t) \\ 
     y(t) &= Cx(t)
\end{align}
\end{subequations}
where $u(t)$,  $x(t)$ and  $y(t)$ are the control law, the state vector and the regulated output respectively,  where
\begin{subequations}\label{eq_etat_mich}
 \begin{align}
     x(t)^{T} &= \begin{bmatrix}
         i_p & i_s & V & v_{c_s} & v_{c_1} & \ldots & v_{c_6}
     \end{bmatrix}\label{eq_statex}\\
     y(t) &= I(t). \label{eq_measy}
\end{align} 
\end{subequations}

The state-space matrices are given by
\begin{align*}
    A \!&= \begin{bmatrix}
        \dfrac{-R_{\ell_p}}{L_p} & 0 & \dfrac{-1}{L_p} & 0& 0_{1\times 6} \\
        0 & \dfrac{-R_{\ell_s}}{L_s} & \dfrac{-1}{L_s} & \dfrac{-1}{L_s} & 0_{1\times 6} \\ 
        \dfrac{1}{C_p} & \dfrac{1}{C_p} & \dfrac{-1}{R_0 C_p} & 0& \dfrac{1_{1\times 6}}{R_0 C_p} \\ 
        0 & \dfrac{1}{C_s} & 0 & 0 & 0_{1\times 6} \\ 
        0_{6\times1} & 0_{6\times1} & \cA_0 & 0_{6\times1} & \cA 
    \end{bmatrix} \\
   B &=  \begin{bmatrix}
        \dfrac{1}{L_p} \\ 
        \dfrac{1}{L_s}\\
        0_{8\times1}
    \end{bmatrix},
     \\ 
    C& = \begin{bmatrix} 0 & 0 & \dfrac{1}{R_0} & 0 & \dfrac{-1_{1\times 6}}{R_0}  \end{bmatrix}
\end{align*}
where $\cA\in{\R}^{6\times 6}$ and $\cA_0\in{\R}^{6\times1}$ are given by (for $h=1,\ldots,6$ and $q=1,\ldots,6$)
\begin{align*}
\cA(h,h)&=\dfrac{-1}{R_h C_h} +\dfrac{-1}{R_0 C_h}, \\
\cA(h,q)&=\dfrac{-1}{R_0 C_h} \qquad h\neq q, \\
\cA_0(h)&=\dfrac{1}{R_0C_h}.
\end{align*}

In this paper, the RES supplied voltage $V_{RES}(t)$ is known and bounded as  
\begin{align}
V_{RES}(t) \in [V_{RES_m} \,\,\, V_{RES_M}]  
\label{eq_VRES}
\end{align}
where $V_{RES_m} = 25\,$V and $V_{RES_M} =55\,$V.

Based on  Faraday's law \cite{AZBPR:25}, the flow rate hydrogen $ \dot{m}_{H_2}$ produced by PEMWE is given by  
\begin{align}
    \dot{m}_{H_2} = \frac{N_c I}{2F} \eta_{H_2}
    \label{eq_debit}
\end{align}
where $N_c =3$ is the number of cells in the stack, $F = 96485$ is Faraday's constant, and $\eta_{H_2}$ is Faraday's efficiency coefficient taken as $0.97$. So, considering the hydrogen flow rate as the control objective is like controlling the current $I$ at the desired setpoint $I^{\star}$.  

\subsection{Problem statement}\label{ssec_problem}
The measured current $y_m$ is expressed as 
\begin{align}\label{eq_fault}
    y_m= I_m = y + f + \omega
\end{align}
where $y$ is defined in \eqref{eq_measy},  $f$ is a sensor fault, and $\omega$ is  a zero-mean white Gaussian noise.

If the sensor fault $f$ is not taken into account in the synthesis of the control law, then it is $I_m$ and not $I$ that converges to the desired setpoint value $I^{\star}$. So an AFTC is therefore necessary to compensate the effect of $f$ on the closed-loop behavior in order to ensure that $I$ converges to $I^{\star}$ despite  $f\neq 0$ and the variations of $V_{RES}(t)$.

This problem is solved in Sect. \ref{sec_SFA} by employing our model-free viewpoint on fault detection, estimation, and accommodation.
\begin{remark}
  It is important to note that the proposed approach detailed in Sect. \ref{sec_SFA} does not use  Eq. \eqref{eq_sys2}, which will be used in Sect. \ref{sec_Sim} to emulate the PEMWE + SIBC behavior.~\hfill~$\square$
\end{remark}

%\color{blue}
\section{MODEL-FREE SENSOR FAULT ACCOMMODATION FOR PEMWE} \label{sec_SFA}
Before presenting the AFTC procedure, we briefly describe the model-free control design that is used in this paper. 
\subsection{Model-free control}\label{ssec-MFC}

Model-free control is based on an \emph{ultra-local} model given by \cite{FlJ:13,FlJ:22}
\begin{equation}%\label{A11}
\dot{y}_m = \cF + \alpha u
\label{ul}
\end{equation}
where $\cF$ corresponds to the unknown part, $u$ and $y_m$ are the control input and measured output, $\alpha$ is an appropriate constant coefficient. An estimation of $\cF$ reads \cite{FlJ:13}
\begin{equation*}\label{integral}
{\small \widehat{\cF}(t)  =-\frac{6}{\tau^3}\int_{t-\tau}^t \left\lbrack (\tau -2\sigma)y_m(\sigma)+\alpha\sigma(\tau -\sigma)u(\sigma) \right\rbrack d\sigma }
\end{equation*}

The corresponding closed-loop control law, or \emph{intelligent proportional controller} (\emph{iP}), reads \cite{FlJ:13} 
\begin{equation}\label{ip}
u = - \frac{\widehat{\cF} - \dot{y}^\star + K_p e}{\alpha}
\end{equation}
where $y^\star = I^\star$ is the reference trajectory, $e = y_m - y^\star$ is the tracking error, while $K_p$ is a tuning gain. %For more detail on Model-free control, the readers can refer to \cite{FlJ:09a}. 

In the presence of a sensor fault $f \neq 0$, the tracking error $e$ converges to 0 but $y$ does not track asymptotically $y^{\star}$ since $y_m \neq y$.

\subsection{Model-free  active fault tolerant control}\label{ssec_AFTC}

\subsubsection{Residual generation}\label{sssec_residual}
We consider the following residual 
\begin{equation}
r(t) = y_{m}(t) - \beta(t) \widehat{y}_m(t)  
\label{eq_residual}
\end{equation}
where $\widehat{y}_m$ is given by 
\begin{equation}
\widehat{y}_m (t) = \int_{0}^{t} \left( \widehat{\cF} (\sigma) + \alpha u(\sigma) \right) \textrm{d}\sigma + \widehat{y}_m (0)
\label{eq_hateym}
\end{equation}
and  $\beta(t)$ is a parameter to be determined. For sensor fault  detection and estimation purposes the residual in Eq. \eqref{eq_residual} must satisfy the following conditions
\begin{itemize}
    \item if $f(t)=0$, then $r(t)=0 $,   $\forall\,\, V_{RES}(t)$, 
    \item if $f(t)\neq0$, then $r(t)\neq0 $,   $\forall\,\, V_{RES}(t)$. 
\end{itemize}

So, when $f(t)=0$, the residual $r(t)$ is equal to 0, if 
\begin{equation}
\beta(t) = \frac{y_m(t)}{\widehat{y}_m(t)} = \frac{y_m(t)}{\int_{0}^{t} \left( \widehat{\cF} (\sigma) + \alpha u(\sigma) \right) \textrm{d}\sigma + \widehat{y}_m (0)}
    \label{eq_beta_y}
\end{equation}
where the steady-state values of the $y_m$ and $\widehat{y}_m$ are used. It has been shown \cite{AJZ:25,AZJP:24c} that $\beta(t)$ in Eq. \eqref{eq_residual} is constant for Linear Time Invariant (LTI) systems.  But the PEMWE system in \eqref{eq_etat_mich} is not LTI due to the variations of $V_{RES}$. Those variations are influencing $\widehat{y}_m(t)$ and $\beta(t)$. It follows that $\beta(t)$ should be expressed as depending on $V_{RES}(t)$. 
Since the iP controller guarantees local stability, the steady-state is still achieved and the control input $u(t)$ in steady-state can be expressed as follows 
\begin{equation}
u(t) = c(t)y(t) = c(t)y_{m}(t)
\label{eq_steady}
\end{equation}

To approximate the integral in Eq. \eqref{eq_beta_y}, the well-known rectangle method is used, where $T_{e}$ is the sampling time: 
\begin{equation*}
\int_{0}^{t} q(\sigma) \textrm{d}\sigma \simeq   \sum_{i=1}^{k} q(i)T_{e}
\label{eq11}
\end{equation*}
and
\begin{equation*}
 \dot{q}(t) \simeq \delta_{q}(i) =  \frac{q(i) - q(i-1)}{T_{e}}
\label{eq10}
\end{equation*}

After straightforward computations, $\widehat{y}_m$ is expressed as follows (see \cite{AZJP:25} for more details):
\begin{equation}
\widehat{y}_m(kT_e) = y_m(kT_e) + \alpha T_{e}u(kT_e) - y_m(0) + \widehat{y}_m(0)
\label{eq18}
\end{equation} 
where $\widehat{\cF}(kT_e)$ is given by (see Eq. \eqref{ul})
\begin{equation}
\widehat{\cF}(kT_{e}) = \delta_{y_m} (kT_{e}) - \alpha u((k-1)T_e)
\label{eq12}
\end{equation}

Insert Eq. \eqref{eq_steady} in Eq. \eqref{eq18}, assume that $y_m(0) \simeq \widehat{y}_m(0)$, then 
\begin{equation}
\beta(kT_e) =  \frac{y_m(kT_e)}{y_m(kT_e) \left( 1  + T_{e} \alpha c(kT_e) \right)}   = \frac{1}{1  + T_{e} \alpha c(kT_e)} 
\label{eq21}
\end{equation}

Let $V_{{RES}_i}$ and $c_i$ be the values of $V_{RES}(k_iT_e)$ and $c(k_iT_e)$ where $(k_iT_e)$ is a given time-instant in the steady-state behavior, where $k> k_i$.

Since the closed-loop stability and steady-state are achieved, $\dot{x} =0$ in Eq. \eqref{eq_sys2}, $c(k T_e)$ in Eq. \eqref{eq_steady} reads 
\begin{equation}
c(k T_e) = c_i(1+\delta_c(kT_e))
\label{eq_c}
\end{equation} 
From 
\begin{equation}
V_{RES}(k T_e) = V_{{RES}_i}(1+ \delta_{V_{RES}}(kT_e))
\label{eq_Vi_res}
\end{equation} 
we obtain
\begin{equation}
\delta_c(k T_e) = \frac{-\delta_{V_{RES}}(k T_e)}{1+\delta_{V_{RES}}(k T_e)}
\label{eq_deltac}
\end{equation} 
Via Eqs. \eqref{eq_c}, \eqref{eq_Vi_res}, \eqref{eq_deltac} and \eqref{eq21}, $\beta(kT_e)$  becomes
    \begin{equation}
        \beta(kT_e) = \frac{1}{1+ \dfrac{V_{RES_i}(1-\beta_i )}{V_{RES}(kT_e)\beta_i }}
        \label{eq_beta_mich}
    \end{equation}
where $\beta_i=\beta(k_iT_e)$.

\subsubsection{Sensor fault estimation and accommodation}\label{sssec_estimation}

If $f\neq 0$ in Eq. \eqref{eq_fault}, we obtain
\begin{align}
u(kT_e) &= c(kT_e)y(kT_e) \notag \\
&= c(kT_e)(y_{m}(kT_e) - f(kT_e) - \omega(kT_e))
\label{eq25} \\
\widehat{y}_m(kT_e)  &= y_{m}(kT_e) + \alpha T_{e}c(kT_e) (y_{m}(kT_e) - \widehat{f}(kT_e))
\label{eq26} 
\end{align}
where the initial condition is chosen by $\widehat{y}_m(0) \simeq y_{m}(0)$, and $\widehat{f}$ is the estimation of the sensor fault $f$ to be determined. Notice that $\widehat{f}$ is used in Eq. \eqref{eq26} in place of $f$ since $\widehat{y}_m$ does not depend explicitly of the noise $\omega$.

Via  Eqs. \eqref{eq_beta_mich}, \eqref{eq26}, and  \eqref{eq_residual}, we get 
\begin{align}\label{eq_f}
   r(kT_e) &= y_{m}(kT_e) -  \frac{1}{1  + T_{e}\alpha c(kT_e)}\times \notag \\
   & \qquad \left(y_{m}(kT_e) (1+\alpha T_{e}c(kT_e)) - \alpha T_{e}c(kT_e) \widehat{f}(kT_e)\right) \notag \\ 
   &= \frac{\alpha T_{e}c(kT_e)\widehat{f}(kT_e)}{1+\alpha T_{e}c(kT_e)} \notag \\ 
   &= \widehat{f}(kT_e)\left( 1 - \frac{1}{1+\alpha T_{e}c(kT_e)} \right)  \notag \\ 
   & = \widehat{f}(kT_e) (1 - \beta(kT_e))  
\end{align}
and the best estimation of  sensor fault $f(kT_e)$ is  given by 
\begin{equation}
\widehat{f}(kT_e) = \frac{r(kT_e)}{1-\beta(kT_e)}
\label{eq30}
\end{equation}
where $\beta(kT_e) \neq 1$ (see Eq. \eqref{eq21}).

%Let $\rho(kT_e)$ be the threshold of the residual $r(kT_e)$. It is given by 

The threshold of the residual $r(kT_e)$ is called $\rho(kT_e)$. It is given by
\begin{align} 
\rho(kT_e) = \widetilde{f} \abs{1-\beta(kT_e)}
\label{eq_th_add}
\end{align} 
where $\widetilde{f}$ is the minimal absolute value of the sensor fault $\widehat{f}(kT_e)$ to be  detected and estimated.

%The determination of the threshold of $r(kT_e)$ is made with respect to the chosen  minimal absolute value $\widehat{f}(kT_e)$ of the sensor fault $f(kT_e)$ to be  detected and estimated. So using Eq. \eqref{eq30}, this threshold $th$ is computed as

%\subsubsection{Sensor fault accommodation}\label{sssec_accommodation}

Once the sensor fault is detected and estimated, i.e, $\abs{r(t)} > \abs{\rho(t)}$,  the sensor fault accommodation is made by adapting the tracking error defined in \eqref{ip} as follows 
\begin{equation}
  e(kT_e) = y_m(kT_e) - y^\star(kT_e) - \widehat{f}(k T_e)
    \label{eq_error_tracking}
\end{equation}

%In this paper, the model described by \eqref{eq_model} is not used to design the control law, but to emulate the behavior of the PEMFC air-feed system. \hfill 

\section{SIMULATIONS RESULTS} \label{sec_Sim}

\begin{figure}[h!]
    \centering
    \includegraphics[width=1\linewidth]{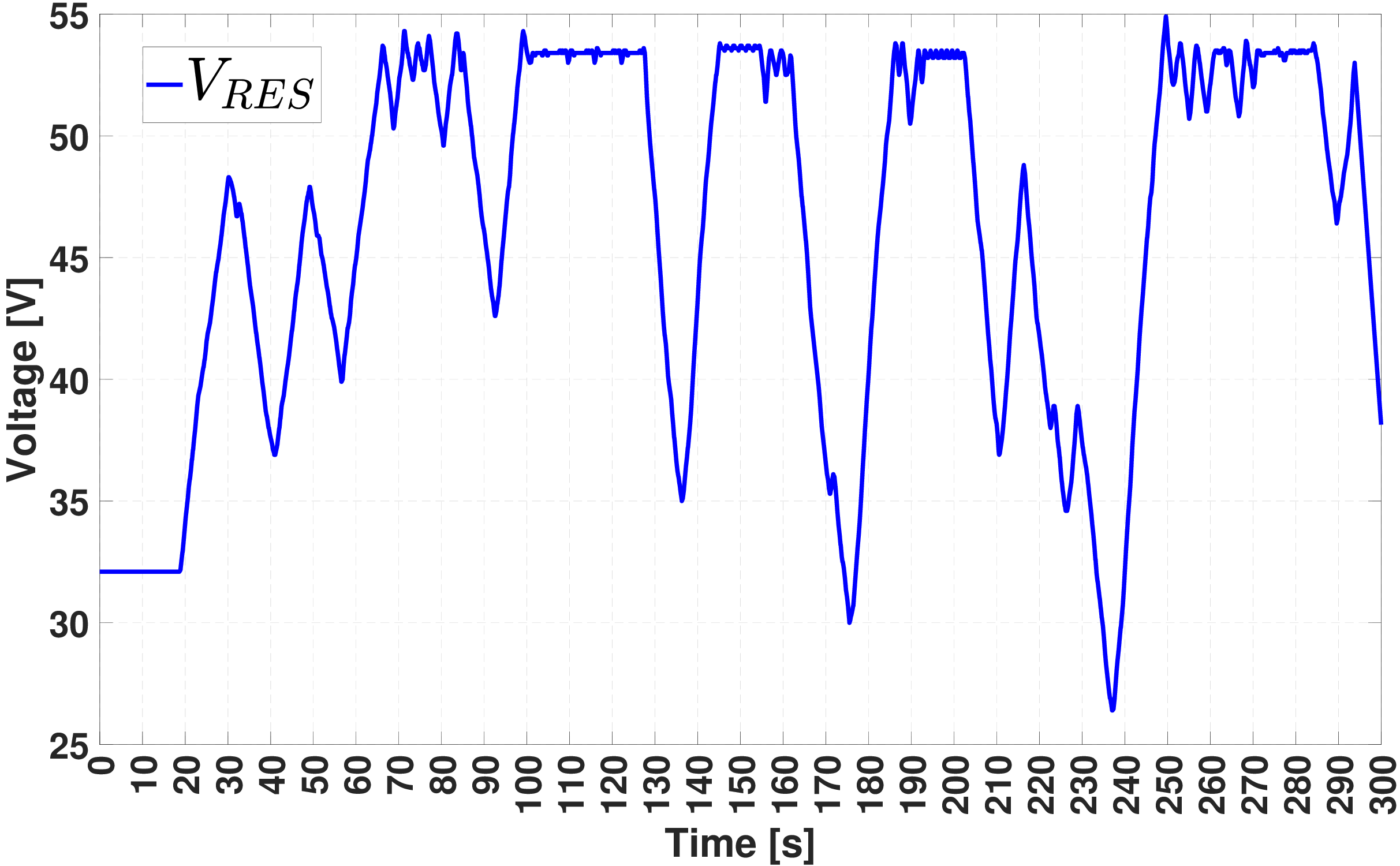}
    \caption{$V_{RES}$ variations}
    \label{fig_VRES_var}
\end{figure}

\begin{figure*}
\centering
\begin{subfigure}{0.4\textwidth}
    \includegraphics[width=\textwidth]{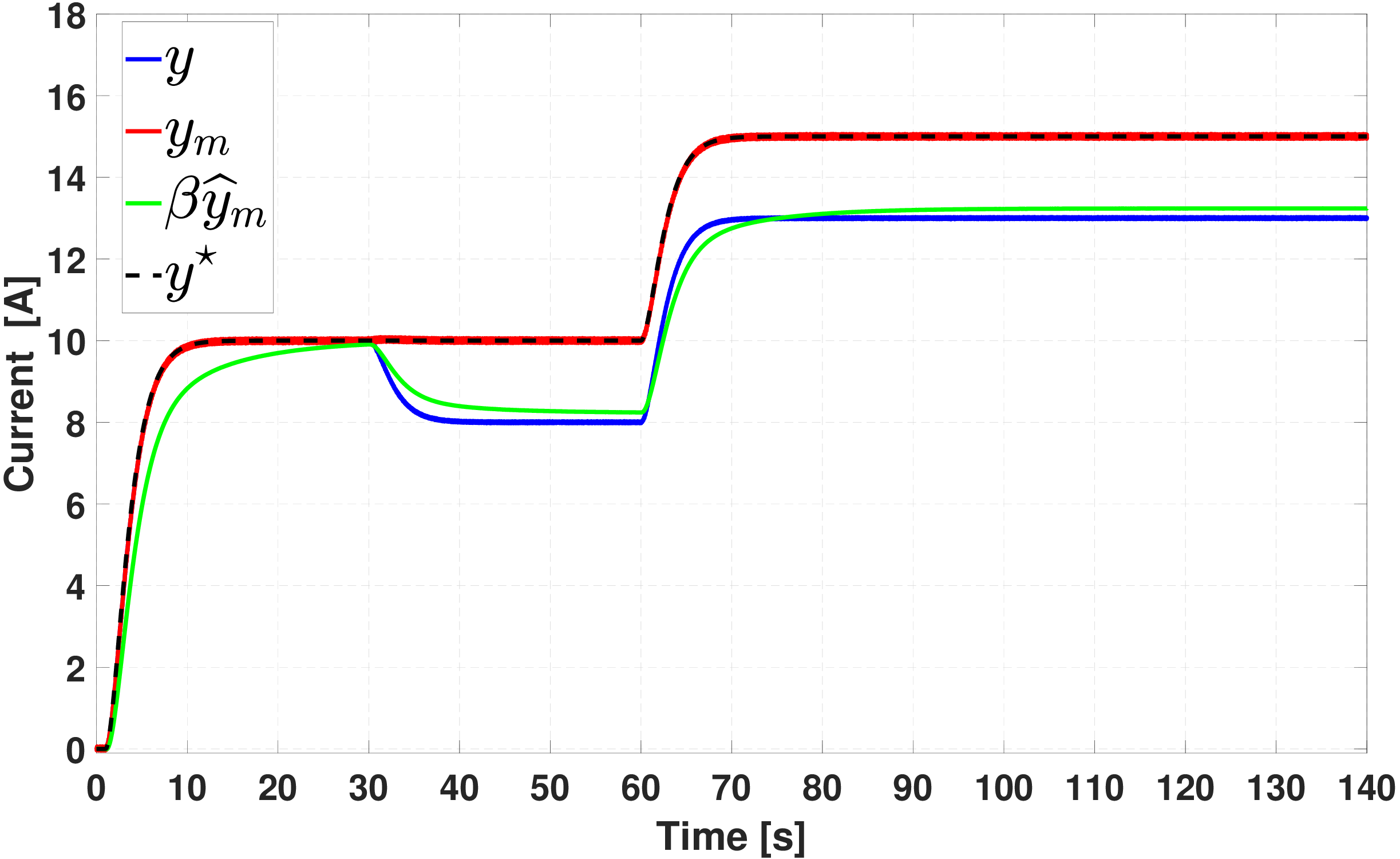}
    \caption{Trajectory tracking without AFTC }
    \label{fig_Vi_fixe_y_sans}
\end{subfigure}
\begin{subfigure}{0.4\textwidth}
    \includegraphics[width=\textwidth]{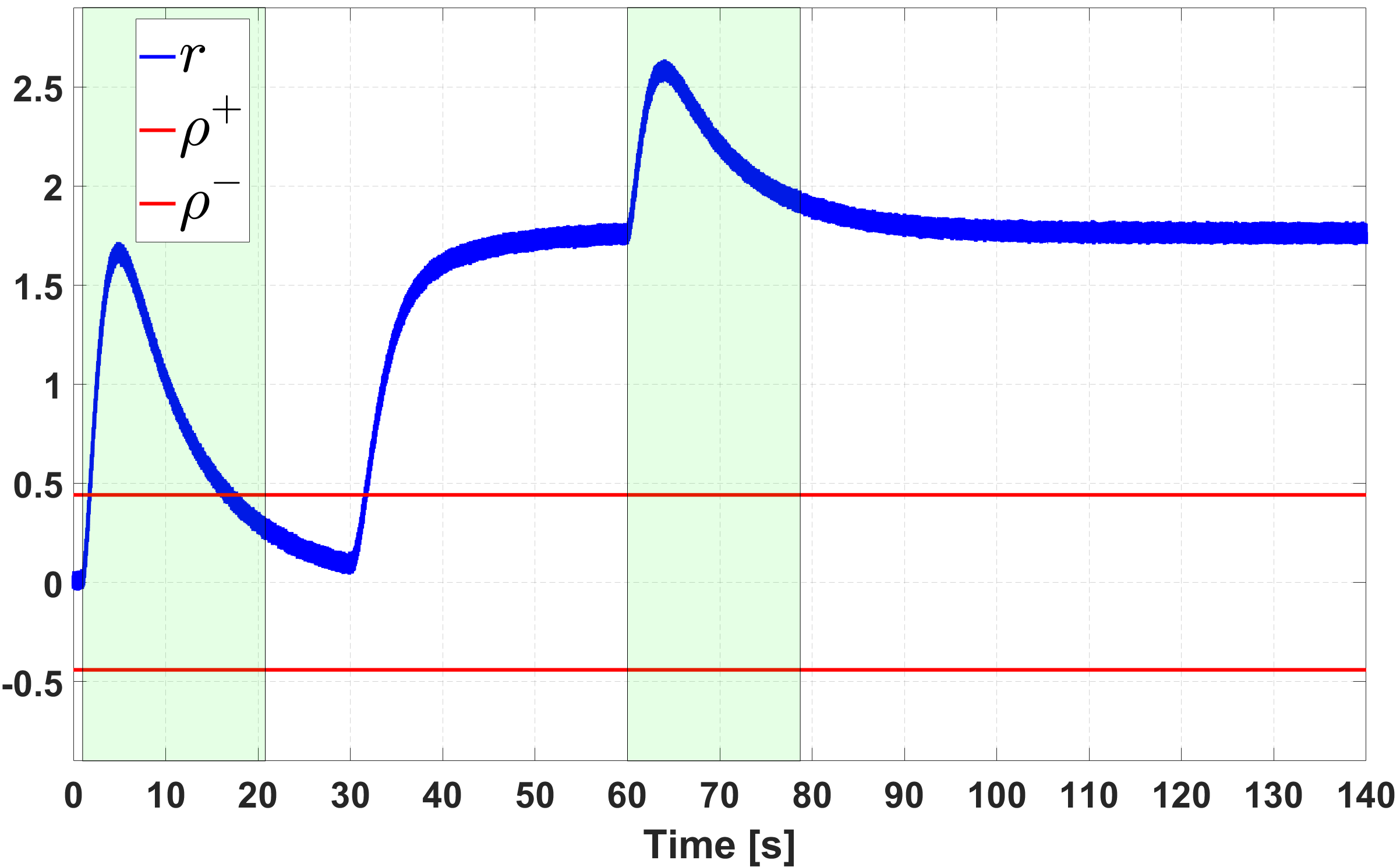}
    \caption{Residual $r$}
    \label{fig_Vi_fixe_r_sans}
\end{subfigure}
\begin{subfigure}{0.4\textwidth}
    \includegraphics[width=\textwidth]{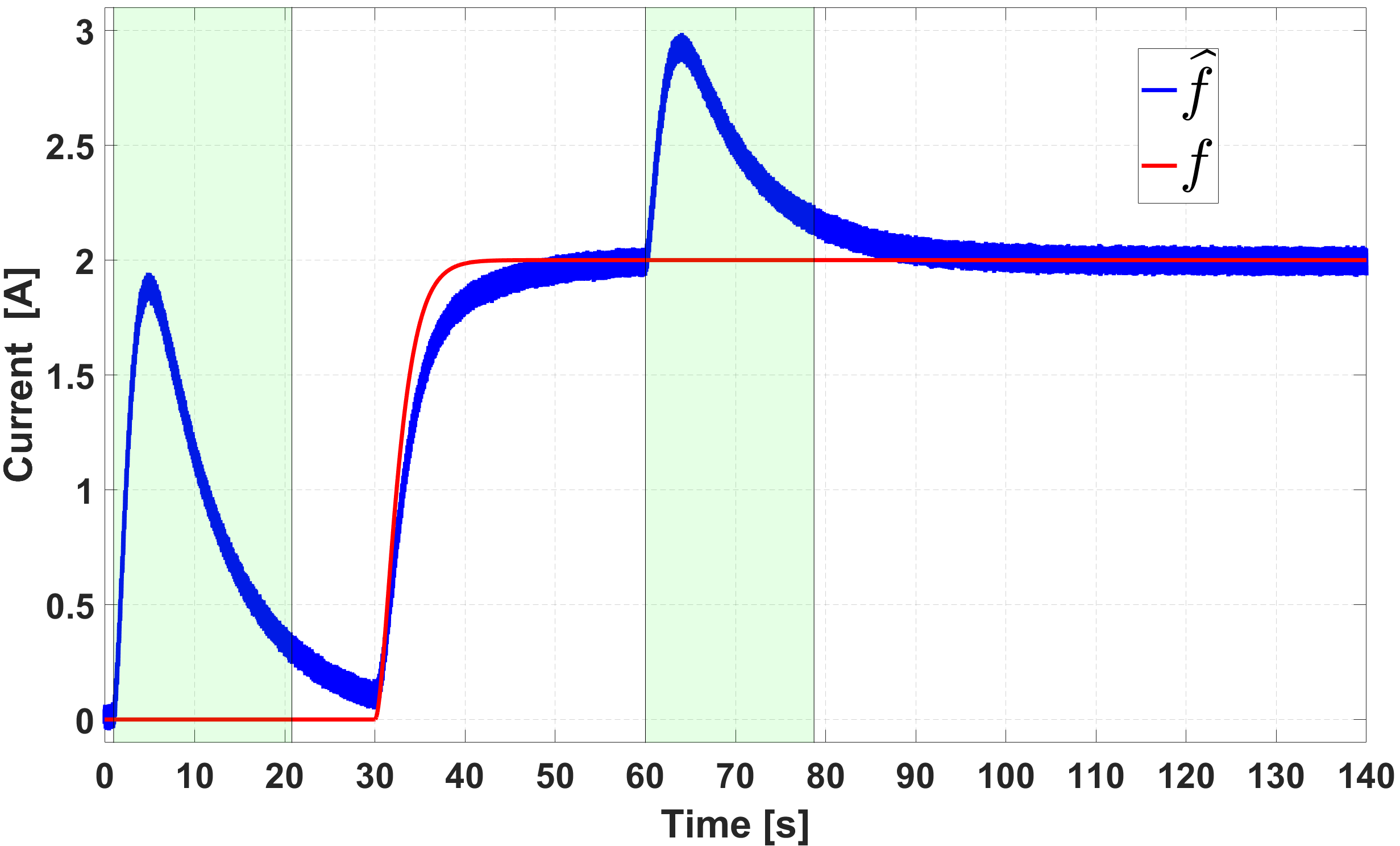}
    \caption{Estimated sensor fault $\widehat{f}$}
    \label{fig_Vi_fixe_f_sans}
\end{subfigure}
\begin{subfigure}{0.4\textwidth}
    \includegraphics[width=\textwidth]{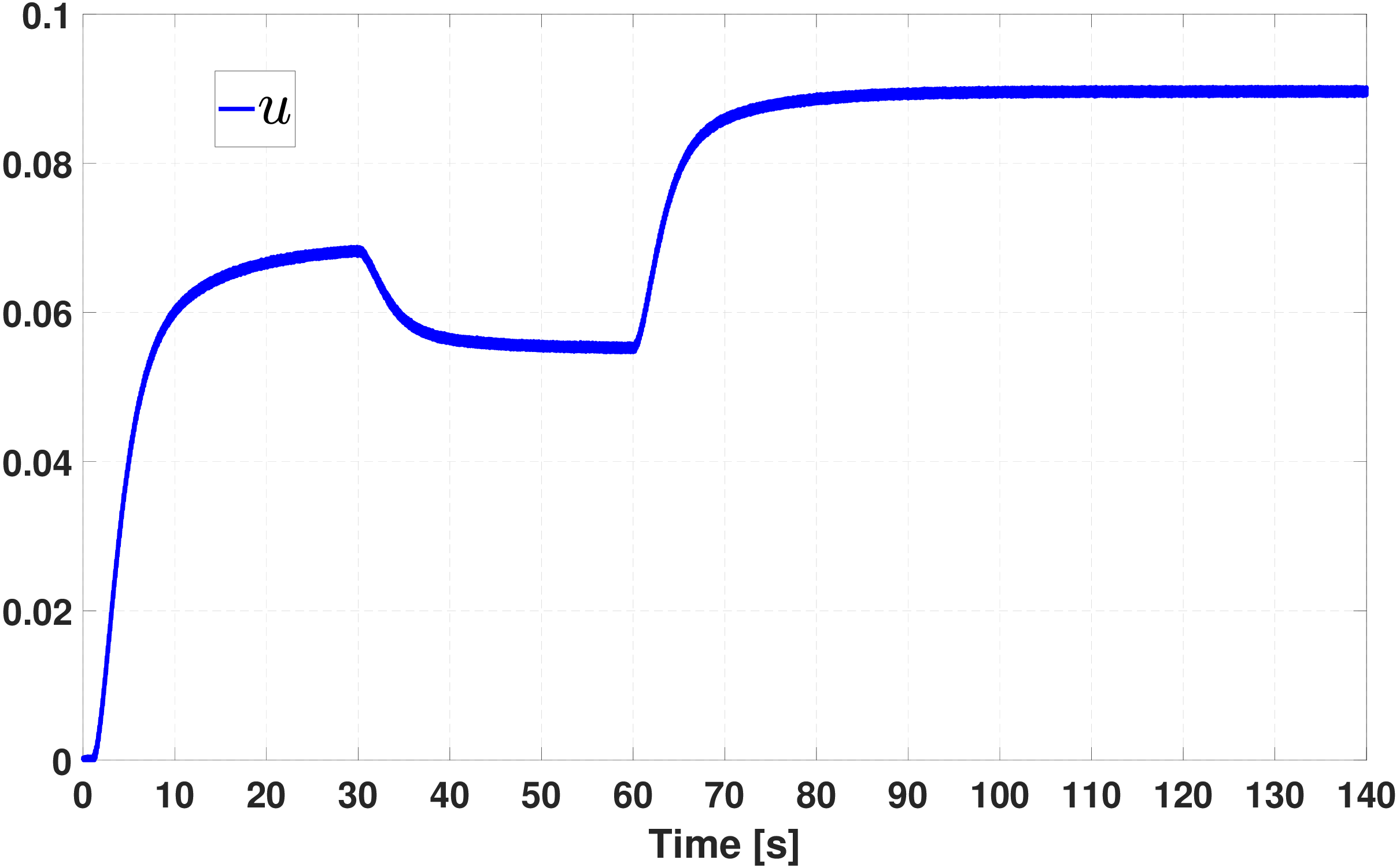}
    \caption{Control input $u$}
    \label{fig_Vi_fixe_u_sans}
\end{subfigure}
\caption{Constant $V_{RES}$ without AFTC}
\label{fig_VRES_fixe_without_ftc}
\end{figure*}

\begin{figure*}
\centering
\begin{subfigure}{0.4\textwidth}
    \includegraphics[width=\textwidth]{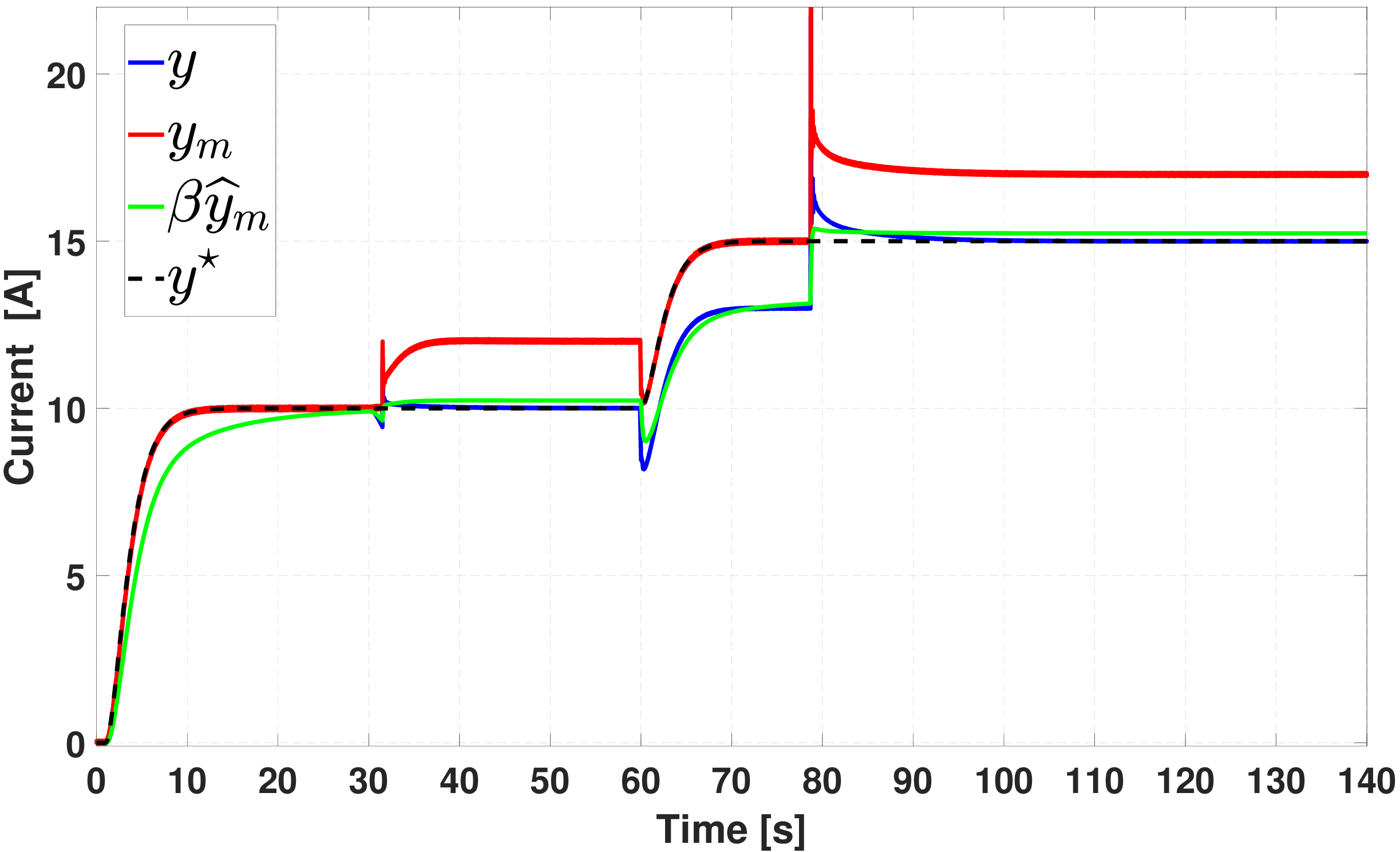}
    \caption{Trajectory tracking with  AFTC}
    \label{fig_Vi_fixe_y_avec}
\end{subfigure}
\begin{subfigure}{0.4\textwidth}
    \includegraphics[width=\textwidth]{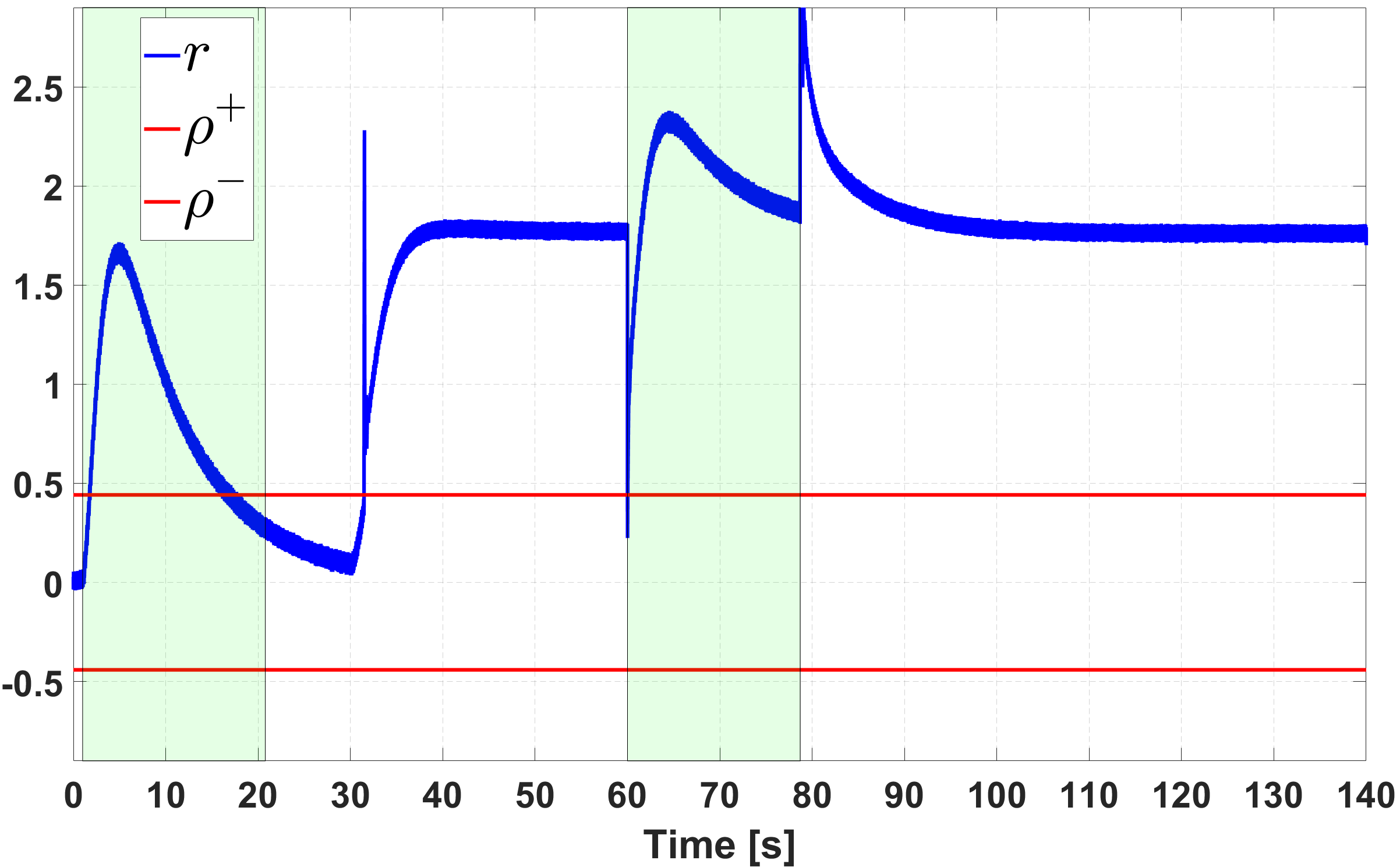}
     \caption{Residual $r$}
    \label{fig_Vi_fixe_r_avec}
\end{subfigure}
\begin{subfigure}{0.4\textwidth}
    \includegraphics[width=\textwidth]{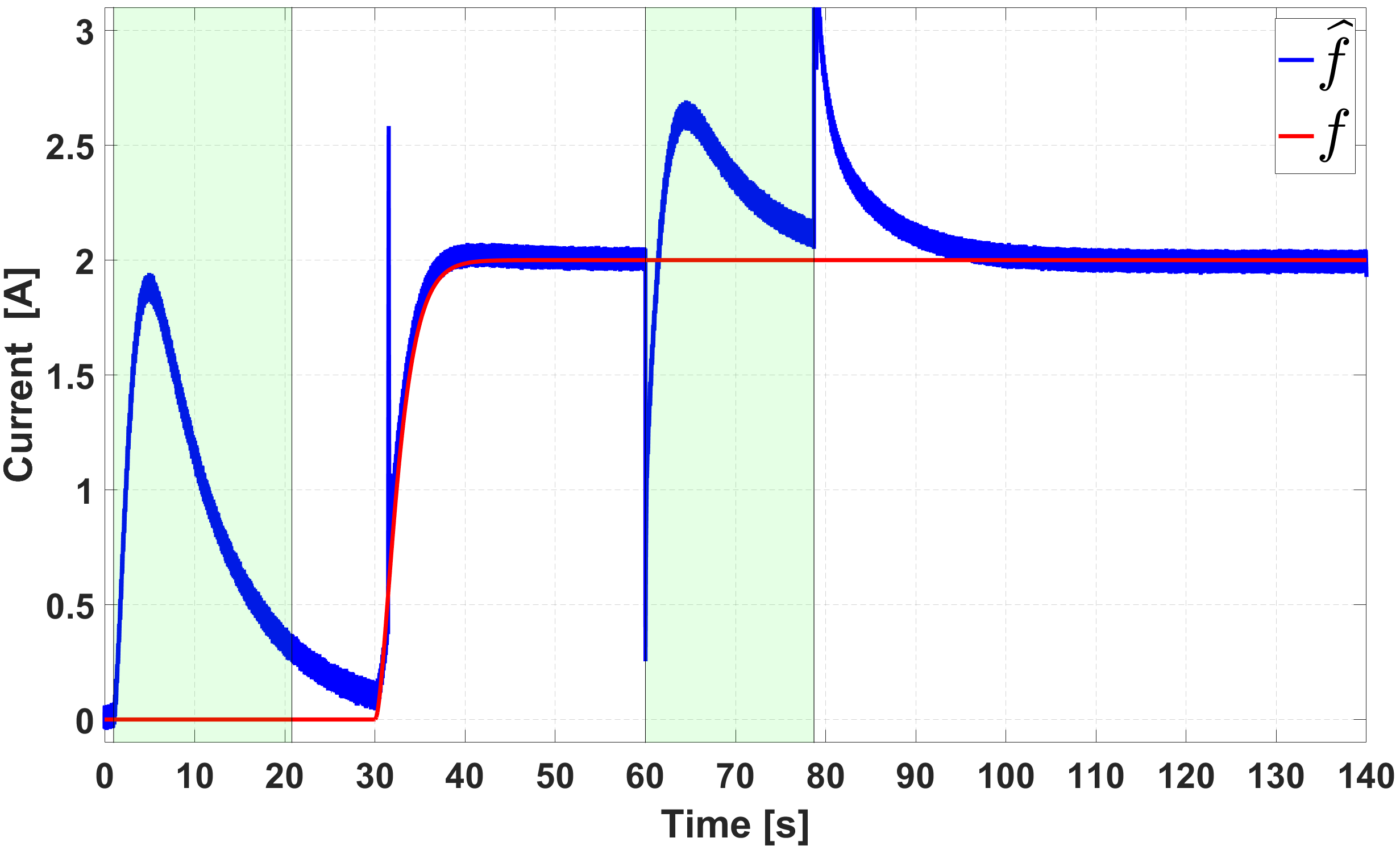}
    \caption{Estimated  sensor fault $\widehat{f}$}
    \label{fig_Vi_fixe_f_avec}
\end{subfigure}
\begin{subfigure}{0.4\textwidth}
    \includegraphics[width=\textwidth]{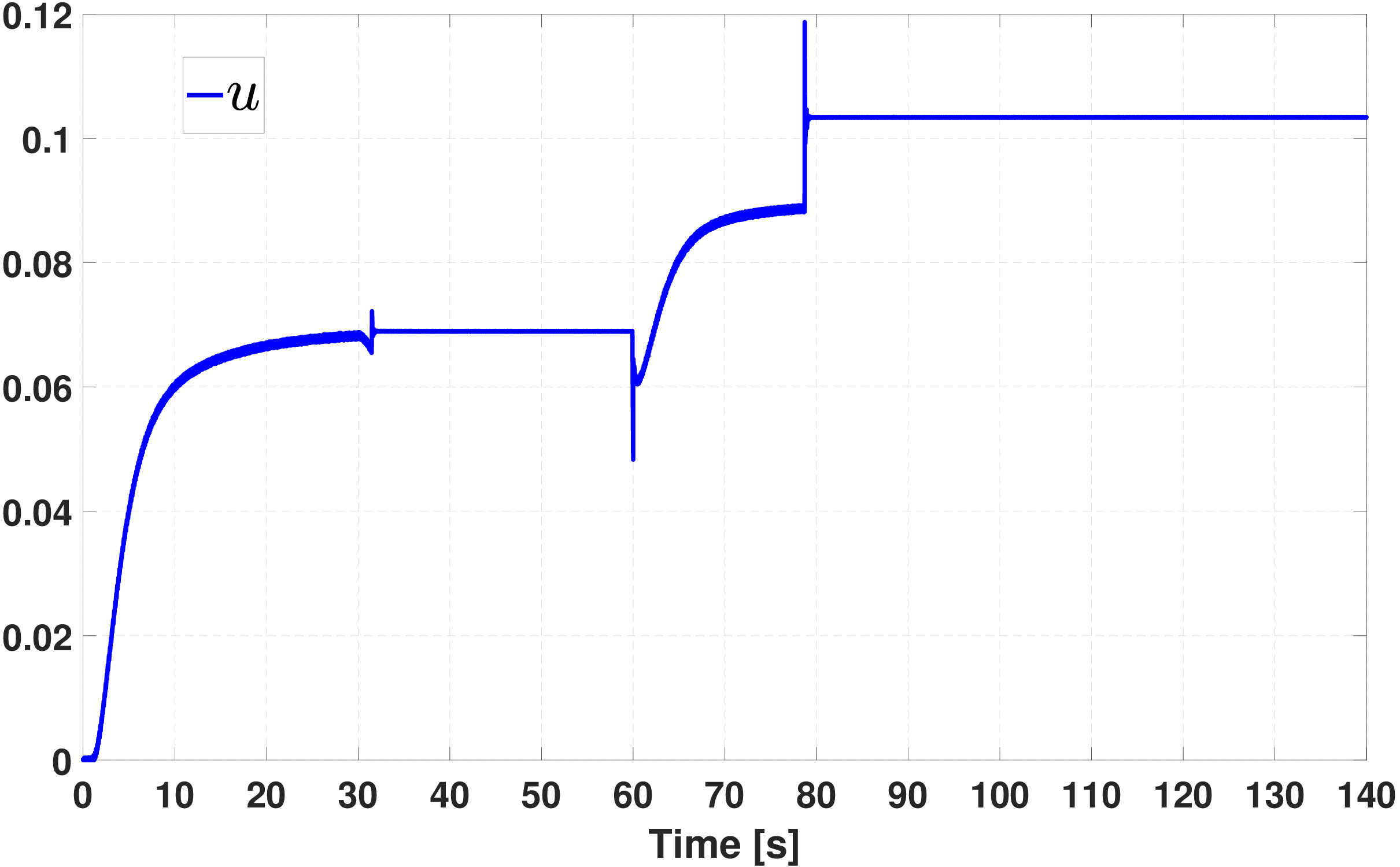}
    \caption{Control input $u$}
    \label{fig_Vi_fixe_u_avec}
\end{subfigure}
\caption{Constant $V_{RES}$ with AFTC}
\label{fig_VRES_fixe_with_ftc}
\end{figure*}

\begin{figure*}
\centering
\begin{subfigure}{0.4\textwidth}
    \includegraphics[width=\textwidth]{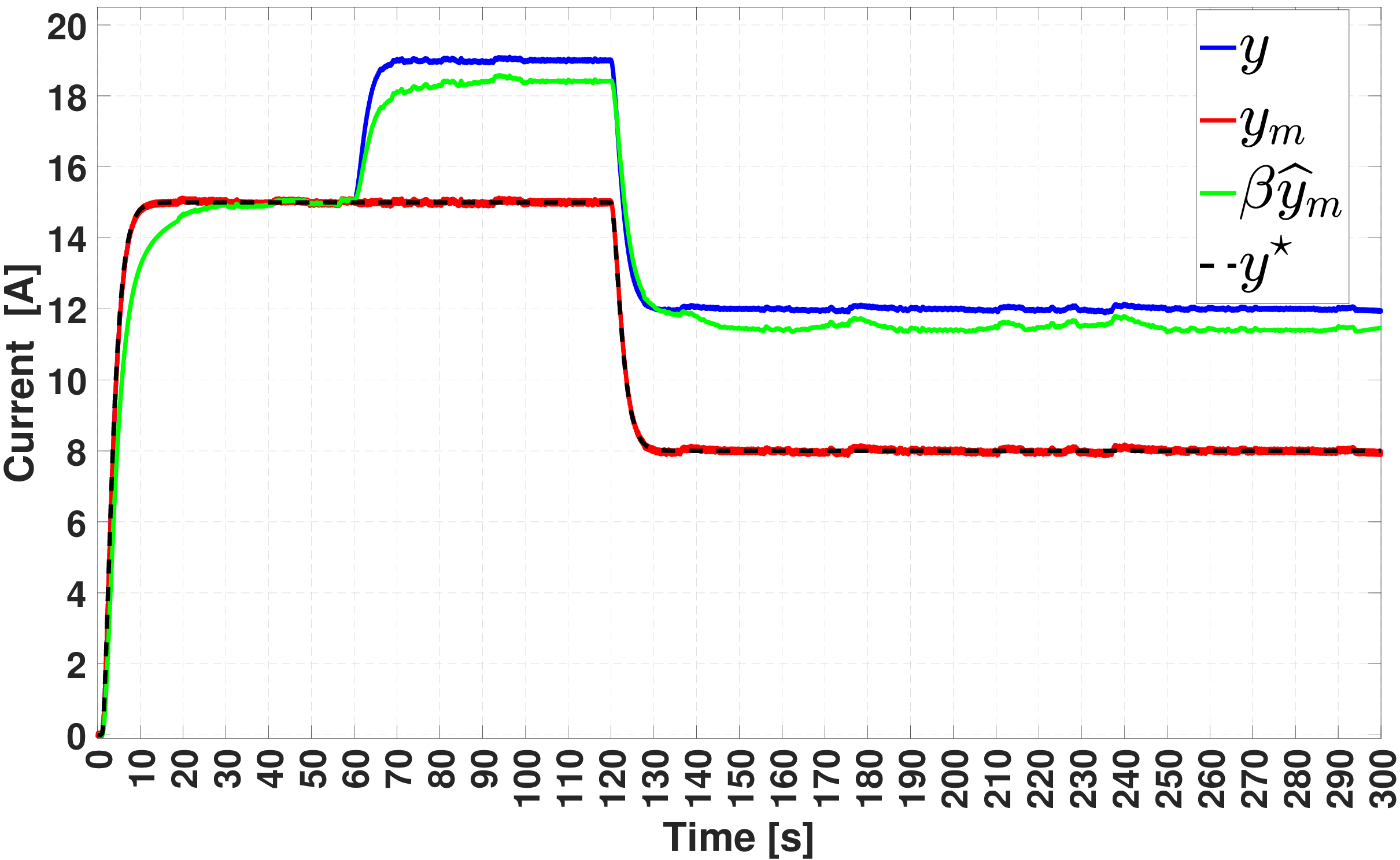}
    \caption{Trajectory tracking without AFTC}
    \label{fig_Vi_var_y_sans}
\end{subfigure}
\begin{subfigure}{0.4\textwidth}
    \includegraphics[width=\textwidth]{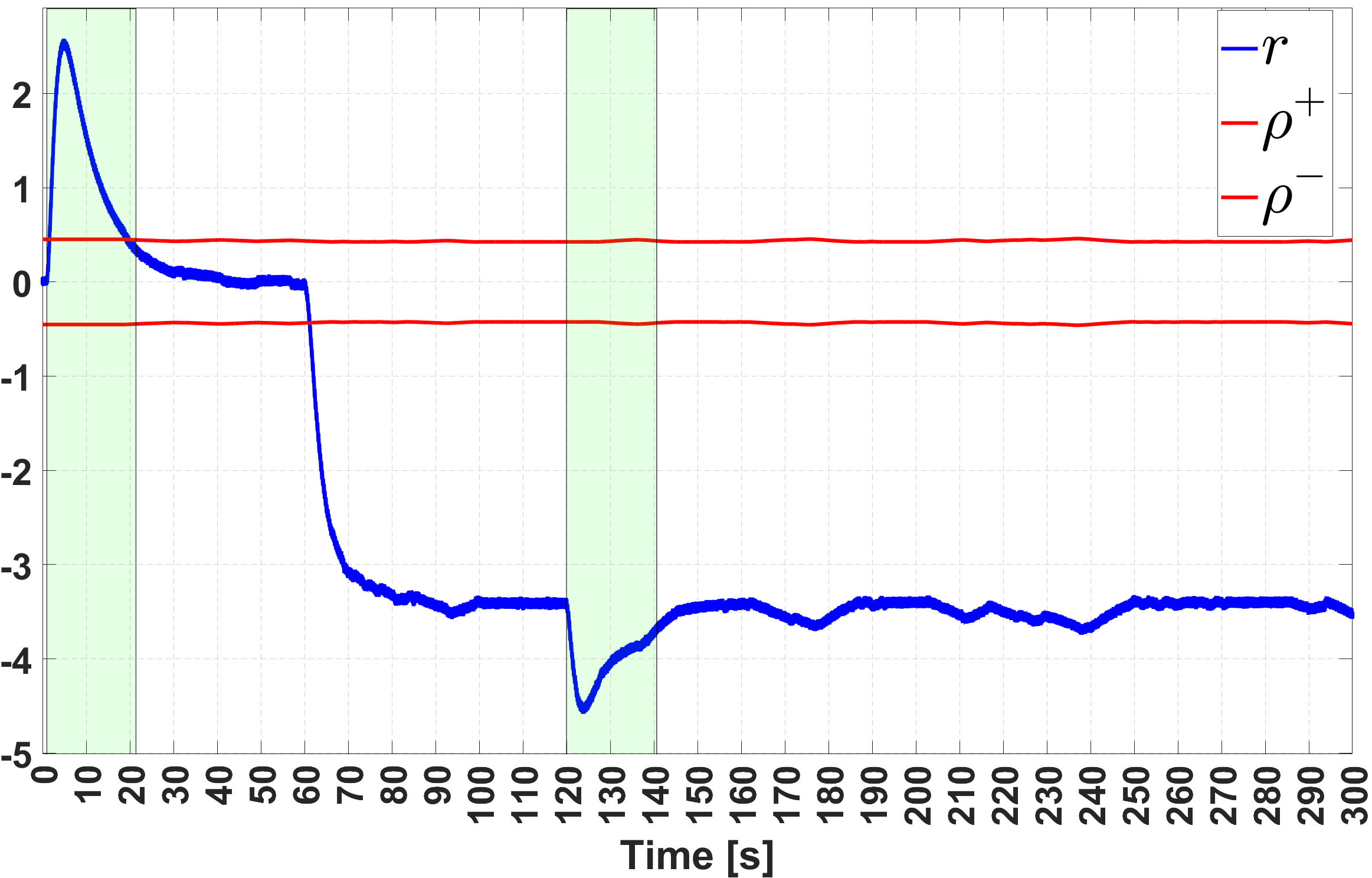}
     \caption{Residual $r$}
    \label{fig_Vi_var_r_sans}
\end{subfigure}
\begin{subfigure}{0.4\textwidth}
    \includegraphics[width=\textwidth]{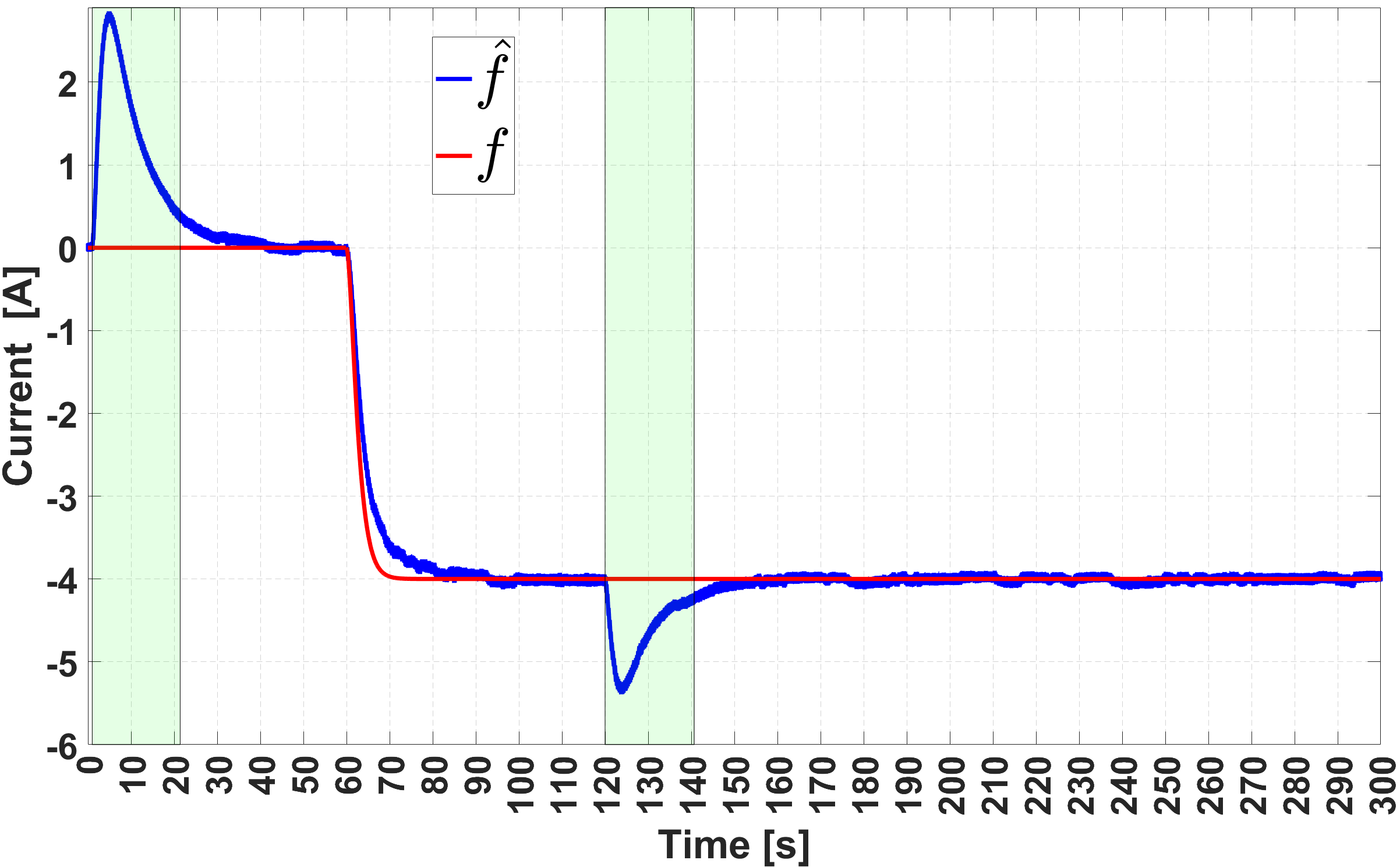}
    \caption{Estimated  sensor fault $\widehat{f}$}
    \label{fig_Vi_var_f_sans}
\end{subfigure}
\begin{subfigure}{0.4\textwidth}
    \includegraphics[width=\textwidth]{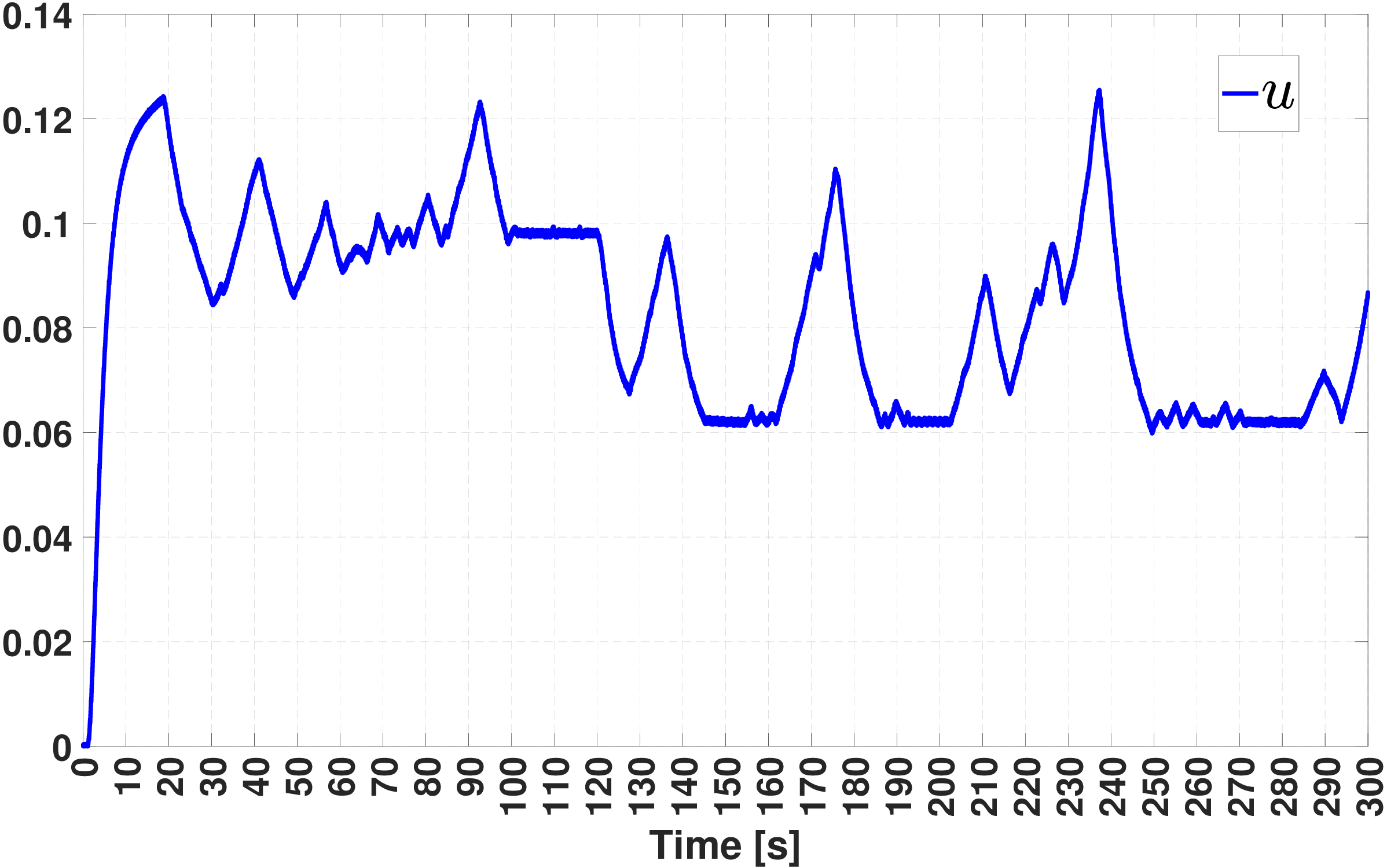}
    \caption{Control input $u$}
    \label{fig_Vi_var_u_sans}
\end{subfigure}
\caption{Variable $V_{RES}$ without AFTC}
\label{fig_VRES_var_without_ftc}
\end{figure*}

\begin{figure*}
\centering
\begin{subfigure}{0.4\textwidth}
    \includegraphics[width=\textwidth]{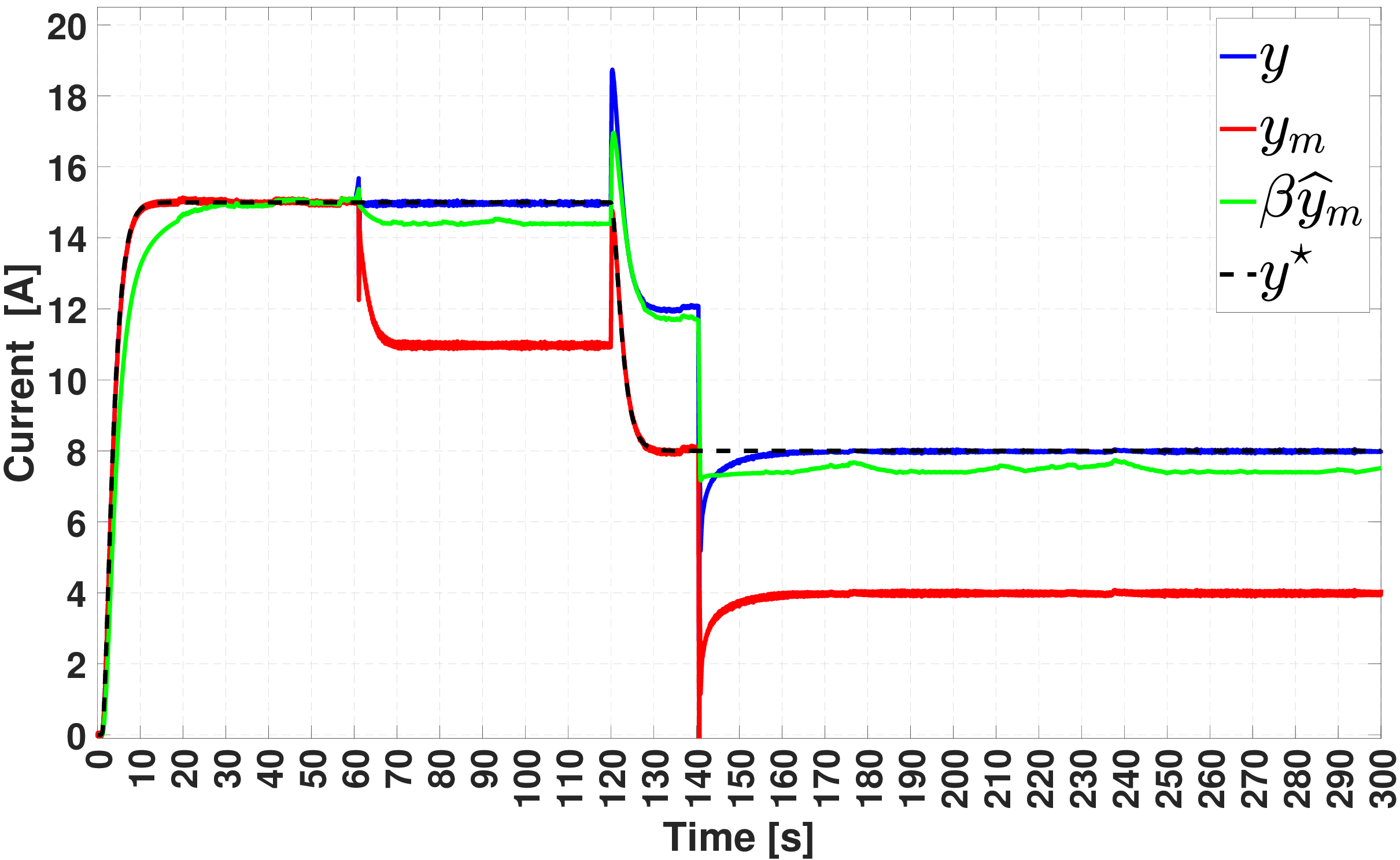}
    \caption{Trajectory tracking without AFTC }
    \label{fig_Vi_var_y_avec}
\end{subfigure}
\begin{subfigure}{0.4\textwidth}
    \includegraphics[width=\textwidth]{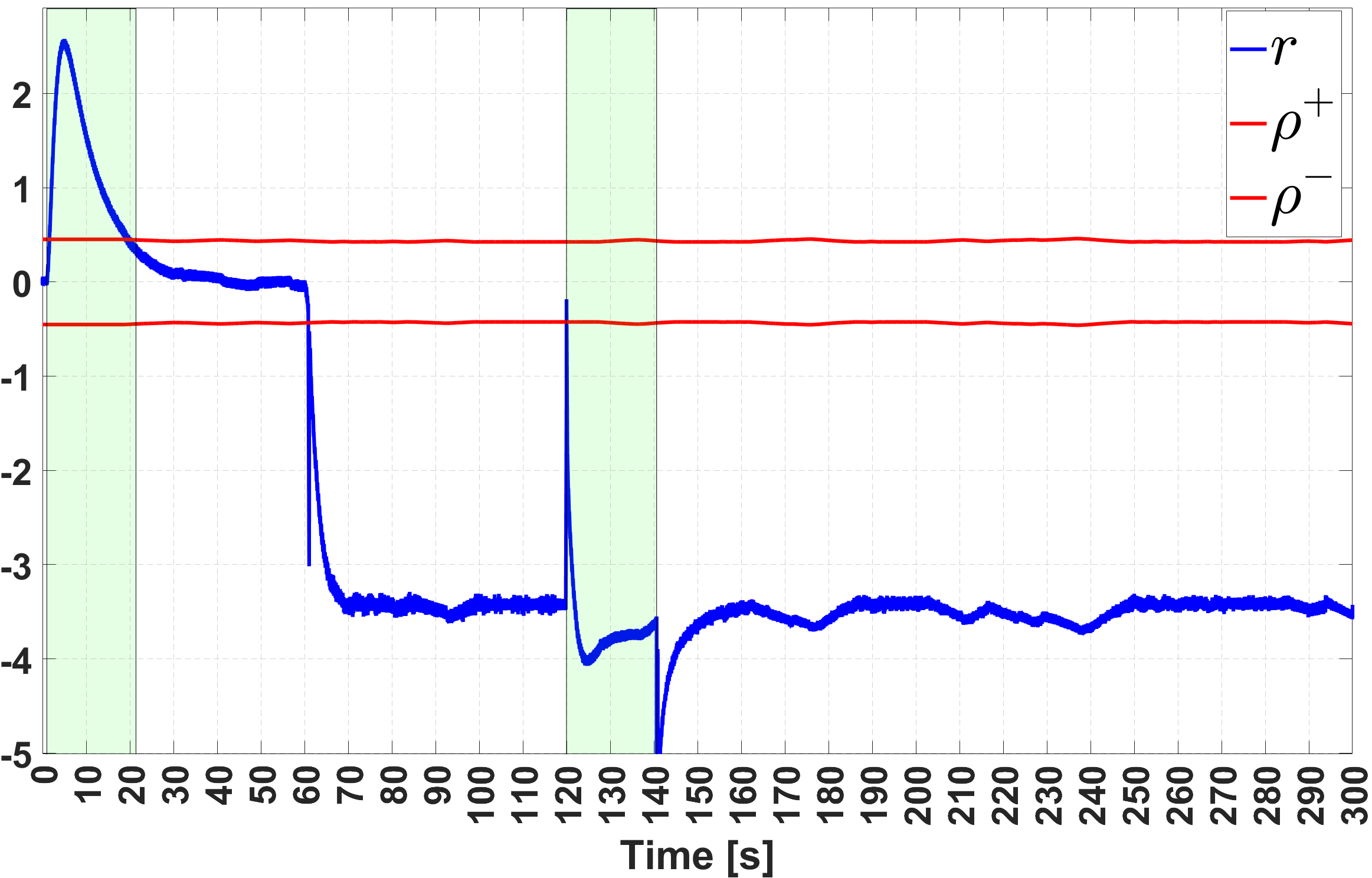}
     \caption{Residual $r$}
    \label{fig_Vi_var_r_avec}
\end{subfigure}
\begin{subfigure}{0.4\textwidth}
    \includegraphics[width=\textwidth]{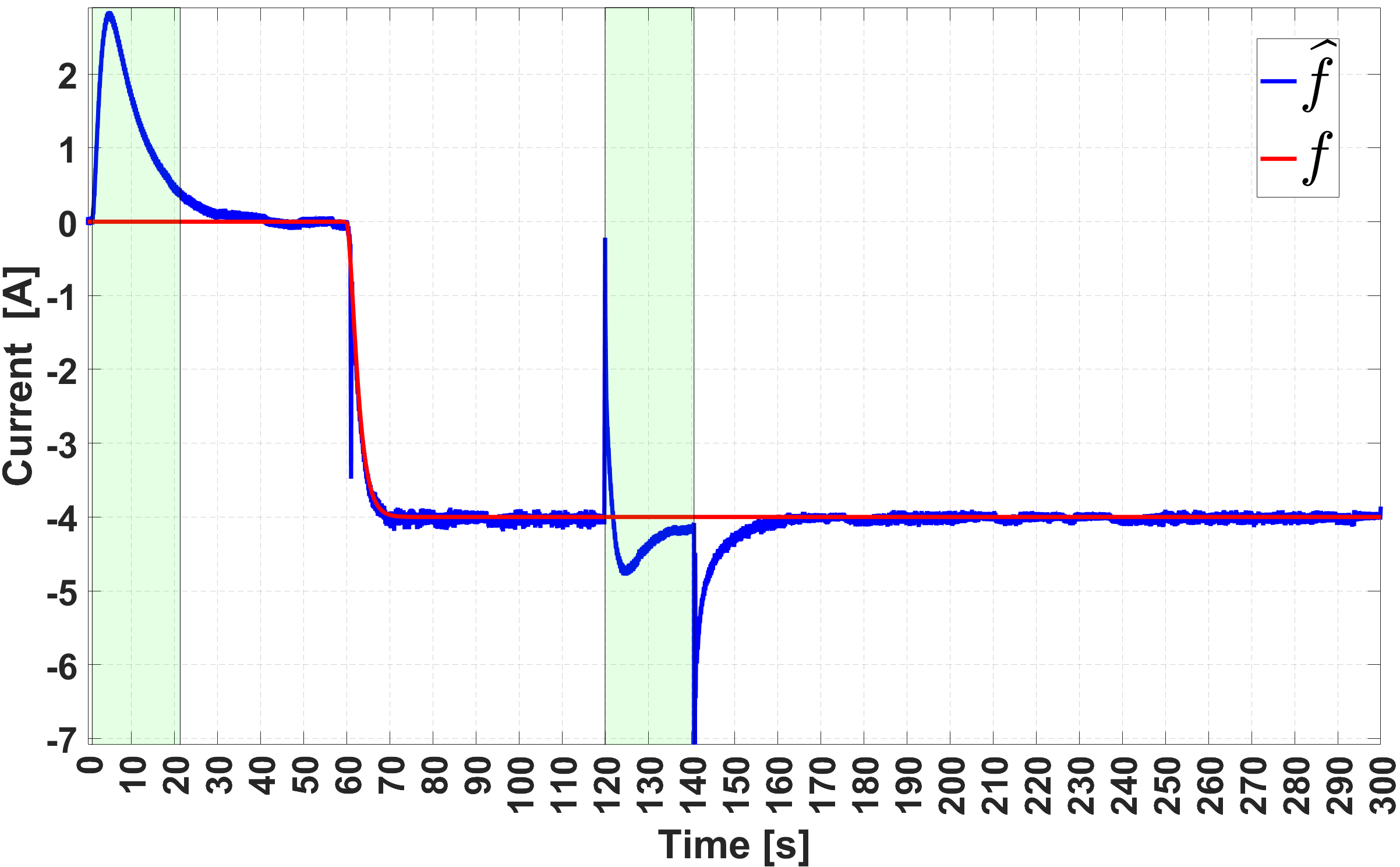}
    \caption{Estimated sensor fault $\widehat{f}$}
    \label{fig_Vi_var_f_avec}
\end{subfigure}
\begin{subfigure}{0.4\textwidth}
    \includegraphics[width=\textwidth]{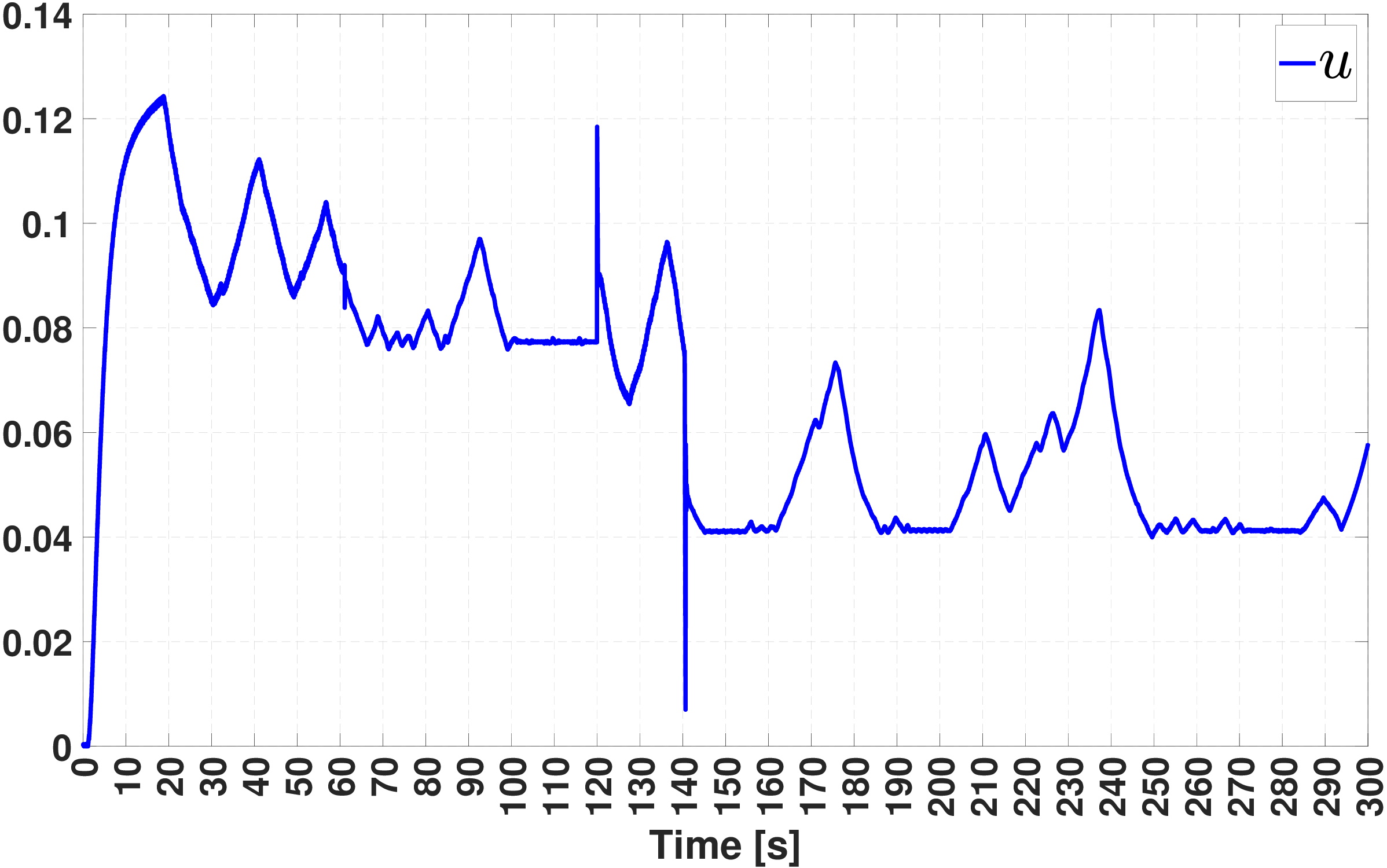}
    \caption{Control input $u$}
    \label{fig_Vi_var_u_avec}
\end{subfigure}
\caption{Variable $V_{RES}$ with AFTC}
\label{fig_VRES_var_with_ftc}
\end{figure*}

Two cases are considered: the \st{1} one with constant $V_{RES} =40\,$V corresponding to Fig. \ref{fig_VRES_fixe_without_ftc} and Fig. \ref{fig_VRES_fixe_with_ftc} and the \nd{2} one with variable $V_{RES}$ which is shown in Fig. \ref{fig_VRES_var} corresponding to Fig. \ref{fig_VRES_var_without_ftc} and Fig. \ref{fig_VRES_var_with_ftc}. In these two cases, there is no sensor fault accommodation procedure in  Fig. \ref{fig_VRES_fixe_without_ftc} and Fig. \ref{fig_VRES_var_without_ftc}, while the AFTC is applied in Fig. \ref{fig_VRES_fixe_with_ftc} and Fig. \ref{fig_VRES_var_with_ftc}. Transient behaviors can be distinguished from steady-state behaviors: colored areas correspond to transient behaviors, while uncolored areas are associated with steady-state behaviors. 

In Fig. \ref{fig_VRES_fixe_without_ftc} to Fig. \ref{fig_VRES_var_with_ftc}, $\rho^{+}(kT_e) = \rho(kT_e)$ and $\rho^{-}(kT_e) = -\rho(kT_e)$ where $\widetilde{f} =0.5\,$A (see Eq. \eqref{eq_th_add}).  In the measurement Eq. \eqref{eq_fault}, $\omega$ is a zero-mean white Gaussian noise with standard deviation $\sigma_\omega =0.0231$. The  value of the \emph{iP} controller parameters are $\alpha = 9750$ and $K_p = 100$. 

For all treated cases, we can see that the \emph{iP} controller ensures closed-loop stability and asymptotic trajectory tracking. So, the AFTC procedure can be applied on the steady-state behavior. For the \st{1} case, the sensor fault $f =2\,$A appears at time $t=30\,$s (see Fig. \ref{fig_Vi_fixe_f_sans} and Fig. \ref{fig_Vi_fixe_f_avec}) and the threshold $\rho$ is constant (see Fig. \ref{fig_Vi_fixe_r_sans} and Fig. \ref{fig_Vi_fixe_r_avec}), while the occurrence of the sensor fault $f =-4\,$A appears at time $t=60\,$s for the \nd{2} case (see Fig. \ref{fig_Vi_var_f_sans} and Fig. \ref{fig_Vi_var_f_avec})  and the threshold $\rho$ is slightly variable (see Fig. \ref{fig_Vi_var_r_sans} and Fig. \ref{fig_Vi_var_r_avec}).

For the case where there is no accommodation procedure, the following comments can be made on Fig. \ref{fig_VRES_fixe_without_ftc} and Fig. \ref{fig_VRES_var_without_ftc}:
\begin{itemize}
    \item the residual $r$ exceeds the threshold $\rho$ at a time-instant very close from the occurrence of the sensor fault, allowing the fault detection and estimation (see Fig. \ref{fig_Vi_fixe_r_sans} and Fig. \ref{fig_Vi_var_r_sans}); 
    \item the sensor fault is well estimated (see Fig. \ref{fig_Vi_fixe_f_sans} and Fig. \ref{fig_Vi_var_f_sans}); 
        \item after the fault occurrence, the measurement $y_m$ follows the desired trajectory $y^\star$ instead of the regulated output $y$ (see Fig. \ref{fig_Vi_fixe_y_sans} and Fig. \ref{fig_Vi_var_y_sans}).
\end{itemize}

For the case where the AFTC procedure is applied, the following comments can be made on Fig. \ref{fig_VRES_fixe_with_ftc} and Fig. \ref{fig_VRES_var_with_ftc}:
\begin{itemize}
    \item the residual $r$ exceeds the threshold $\rho$ at a time-instant very close from the occurrence of the sensor fault, allowing the fault detection and estimation (see Fig. \ref{fig_Vi_fixe_r_avec} and Fig. \ref{fig_Vi_var_r_avec}). At this time-instant, the AFTC procedure is triggered; 
    \item the sensor fault is well estimated (see Fig. \ref{fig_Vi_fixe_f_avec} and Fig. \ref{fig_Vi_var_f_avec});  
        \item once the AFTC procedure is triggered, the regulated output $y$ follows the desired trajectory $y^\star$, which means that the AFTC is effective  (see Fig. \ref{fig_Vi_fixe_y_avec} and Fig. \ref{fig_Vi_var_y_avec}). Notice that, the  AFTC is not triggered in the transient behavior, so  the measurement $y_m$ follows the desired trajectory $y^\star$ instead of the regulated output $y$; 
        \item by comparing Fig. \ref{fig_Vi_fixe_u_sans} and Fig. \ref{fig_Vi_fixe_u_avec} on the one hand, and Fig. \ref{fig_Vi_var_u_sans} and Fig. \ref{fig_Vi_var_u_avec} on the other, we can see how  the AFTC procedure has modified the control input $u$.
\end{itemize}

\section{CONCLUSIONS}\label{sec_Conc}
Detection and accommodation of current sensor fault for the PEMWE were discussed. This measurement is essential for controlling hydrogen production via a water electrolyzer. The method developed does not rely on an analytical model of the process: the detection and compensation of sensor faults are {\em model-free}. The simulation results confirm that the actual current of the electrolyzer is regulated to the desired current despite significant variations in the voltage supplying the process in the presence of a sensor fault. 

The fact that our approach is { \em model-free} and based on an { \em ultra-local} model makes it a good candidate for real-time application on a test bench, with low computational burden. The implementation of this method on a real test bench is being considered as part of future work.

%The detection and accommodation of current sensor fault for PEMWE is addressed in this paper. This measurement is essential for controlling the production of hydrogen via water electrolyzer. The developed method does not relies on the analytical process model: sensor fault detection and accommodation is {\em model-free} which makes it a good candidate for real-time application on test bench with low computational burden. Simulations results confirm that the real current of the electrolyzer is regulated to the desired current under large variations of the voltage supplying the process in the presence of sensor fault. 

%The fact that our approach is {\em model-free} and relies on an {\em ultra-local} model makes it a good candidate for real-time application on test bench with low computational burden. The implementation of this method on actual test bench is envisaged in future work. 

\bibliography{MZ}

\end{document}